\documentclass[superscriptaddress,nofootinbib,aps,prd,preprintnumbers,noshowpacs]{revtex4}
\usepackage{graphicx}
\usepackage{amsmath,amssymb}
\usepackage{color}
\usepackage{hyperref}
\usepackage{slashed}
\usepackage{braket}
\usepackage[utf8]{inputenc}

\definecolor{darkblue}{rgb}{0,0,0.9}
\hypersetup{
	linktocpage,
	colorlinks,
	citecolor=darkblue,
	filecolor=darkblue,
	linkcolor=darkblue,
	urlcolor=darkblue,
	pdftitle={MDM-diphoton},
	pdfauthor={Kirtimaan Mohan}
}

\newcommand{\lag}{\mathcal{L}}
\newcommand{\oli}{\overline}
\newcommand{\al}[1]{\begin{align}#1\end{align}}
\newcommand{\paren}[1]{\left(#1\right)}

\newcommand{\nn}{\nonumber\\}
\newcommand{\bb}{\begin{bmatrix}}
\newcommand{\eb}{\end{bmatrix}}

\newcommand{\cL}[1]{\cos^{#1}\theta_L}
\newcommand{\sL}[1]{\sin^{#1}\theta_L}
\newcommand{\cS}[1]{\cos^{#1}\theta_S}
\newcommand{\sS}[1]{\sin^{#1}\theta_S}

\begin{document}

\title{Minimal Dilaton Model and the Diphoton Excess}
\author{Bakul Agarwal}
\author{Joshua Isaacson}
\author{Kirtimaan~A.~Mohan}
\affiliation{Department of Physics and Astronomy, Michigan State University, East Lansing, Michigan 48824, USA}

\begin{abstract}
In light of the recent 750 GeV diphoton excesses reported by the ATLAS and CMS collaborations, we investigate the possibility of explaining this excess using the 
Minimal Dilaton Model. 
We find that this model is able to explain the observed excess with the presence of additional top partner(s), with same charge as the top quark, but with mass in the TeV region. 
First, we constrain model parameters using in addition to the 750 GeV diphoton signal strength, 
precision electroweak tests, single top production measurements, as well as Higgs signal strength data collected in the earlier runs of the LHC. 
In addition we discuss interesting phenomenolgy that could arise in this model, relevant for future runs of the LHC.
\end{abstract}

\maketitle

\section{Introduction}
Recently both ATLAS~\cite{ATLAS-CONF-2015-081, Moriond-ATLAS} and CMS~\cite{CMS-PAS-EXO-15-004,CMS-PAS-EXO-16-018,Moriond-CMS} reported small excesses in their search for diphoton resonances.
If indeed a resonance exists with a mass of $\sim 750$ GeV, then it certainly must belong to Beyond the Standard Model (BSM) Physics.
There have been several papers investigating a plethora of possible  BSM scenarios~\cite{1512.04850,1512.04913,1512.04917,1512.04921,1512.04924,
1512.04928,1512.04929,1512.04931,1512.04933,1512.04939,1512.05326,1512.06028,1512.05327,1512.05328,1512.05330,1512.05332,1512.05333,1512.05334,1512.05439,1512.05542,1512.05564,1512.05585,
1512.05617,1512.05618,1512.05623,1512.05700,1512.05723,1512.05738,1512.05751,1512.05767,1512.05771,1512.05776,1512.05777,1512.05778,1512.05779,1512.05961,1512.06083,1512.06091,
1512.06106,1512.06107,1512.06113,1512.06297,1512.06335,1512.06376,1512.06426,1512.06670,1512.06508,1512.06560,1512.06562,1512.06587,1512.06671,1512.06674,
1512.06696,1512.06708,1512.06715,1512.06728,1512.06732,1512.06741,1512.06773,1512.06782,1512.06787,1512.06797,1512.06799,1512.06824,1512.06827,1512.06828,1512.06833,1512.06842,
1512.06878,1512.06976,1512.07165,1512.07212,1512.07225,1512.07229,1512.07242,1512.07243,1512.07268,1512.07462,1512.07468,1512.07497,1512.07527,1512.07616,1512.07624,1512.07541,
1512.08507,1512.07645,1512.07672,1512.07733,1512.07789,1512.07853,1512.07885,1512.07889,1512.07895,1512.07904,1512.07992,1512.08117,1512.08184,1512.08255,1512.08307,1512.08323,
1512.08378,1512.08392,1512.08434,1512.08440,1512.08441,1512.08467,1512.08478,1512.08484,1512.08497,1512.08500,1512.08502,1512.08508,1512.09129,1512.09089,1512.08777,
1512.09136,1512.09127,1512.09092,1512.09053,1512.09048,1512.08992,1512.08984,1512.08963,1512.08895,1512.09202,1601.03604,1601.03696,1601.03153,1601.02714,1601.02447,1601.02570,
1601.02490,1601.01712,1601.01676,1601.01828,1601.01569,1601.01571,1601.01144,1601.00661,1601.00836,1601.00638,1601.00624,1601.00618,1601.00602,1601.00534,1601.00285,1601.00586,
1601.00386,1601.01355,1601.00006,1601.03267,1601.02457,1601.02004,1601.01381,1601.00952,1601.00866,1601.00640,1601.00633,1601.04678,1601.00240,1601.04291,1601.04516,1601.04751,
1601.04954,1601.05038,1601.05357,1601.05729,1601.06374,1601.06394,1601.06761,1601.07167,1601.07187,1601.07242,1601.07339,1601.07385,1601.07330,1601.07396,1601.07508,1601.07564,
1601.07774,1602.00004,1602.00475,1602.00949,1602.00977,1602.01092,1602.01377,1602.01460,1602.01801,1602.02380,1602.02793,1602.03344,1602.03604,1602.03607,1602.03653,1602.03877,
1602.04170,1602.04204,1602.04692,1602.04822,1602.04838,1602.05539,1602.05581,1602.05588,1602.07866,1602.08092,1602.09099,1603.03421}.

Interestingly, the diphoton excess is accompanied by an absence of evidence for any excesses in other channels ($hh$, $WW$, $ZZ$, $t\bar{t}$, etc.). This, along with the fact that its mass is much heavier than any of the SM particles, suggests that the $750$ GeV resonance must not couple strongly to the Standard Model (SM) sector. 
Further, data suggests that the width of the new resonance to be not too large. These characteristics imply that the $750$ GeV resonance is most likely the 
lightest BSM particle and that there must be additional heavier charged BSM particles that contribute to loop induced decays of the $750$ GeV resonance.

Motivated by the simple observations made above, about the nature of the resonance, we carry out investigations on the effective Minimal Dilaton 
Model~\cite{Abe:2012eu,Cao:2013cfa}. 
The model consists of vector-like fermionic top partner(s) with mass $M_i$ that characterize the mass gap of dynamical 
symmetry breaking of an approximate scale invariance. The quantum numbers of the top partner(s) are chosen to be identical to that of a right handed top quark. 
This choice is motivated by topcolor~\cite{Hill:1991at} and top seesaw models~\cite{Chivukula:1998wd} that predict a naturally large top quark mass. 
The dilaton field $S$ couples directly only to the Higgs boson and top partner(s). 
Coupling of the dilaton to photons and gluons proceeds only through loops of the top partner(s). Thus the model can predict the production of a dilaton $S$ of 
mass $m_s\sim750$ GeV and its subsequent decay to a pair of photons. 
In this paper we investigate the consistency of the above statement with data.
A similar study was carried out in Ref.~\cite{1601.02570}. Here we extend their analysis to the case where we have non-trivial mixing between the top and its partner(s).

The paper is organized as follows. In the next section, we introduce the model and its parameters. We then discuss various constraints on the model imposed by precision
electroweak tests, the observed Higgs signal strengths, and the single top production rate measurement. 
Following which we study constraints on the parameter space that arises in order to explain the $750$ GeV excess. We further discuss phenomenological implications of the model for future LHC runs before concluding.

\section{Lagrangian and Model Paramaters}
The dilaton is a pseudo-Nambu-Goldstone boson associated with conformal symmetry breaking and couples to the trace of the energy momentum tensor of 
the SM~\cite{Goldberger:2008zz,Foot:2007iy}. It is useful to write down an effective low energy lagrangian using an effective linearized dilaton field $S$,
which is a gauge singlet, as follows~\cite{Abe:2012eu}:
\begin{eqnarray}
\lag = \lag_{\rm SM}
- {1 \over 2 }
\partial_\mu S \partial^\mu S
- \sum_{i=1}^{N_T}\oli{ T_i}\left(\slashed{D}+{M_i \over f} S\right)T_i
- \sum_{i=1}^{N_T}\left[
 y_i' \oli{T_i}_R(q_{3L}\cdot \tilde{H}^c)
+ \text{h.c.} \right]
- \tilde V(S, H).
\label{eqn:lin-lag}
\end{eqnarray}
Here, $\lag_{SM}$ is the SM Lagrangian modulo the Higgs potential, $f=\braket{S}$ is the vacuum expectation value of the dilaton and $T_i$ correspond to the $N_{T}$ number of 
fermionic top partners. Note that the assumption here is that only the Higgs and the top partner(s) $T$ are assumed to couple to BSM dynamics, while the 
remaining SM particles are considered to be spectators to the BSM sector. As a result, the dilaton $S$ couples only to $T_i$ and $H$ fields\footnote{ In typical dilaton models, the entire 
SM particle spectrum is assumed to couple to the dilaton~\cite{Foot:2007iy}.} and does not couple directly to $W$'s, $Z$'s and other fermions of the SM, albeit through possible mixing between the Higgs and dilaton. This mixing 
has been parametrized as follows:
\begin{align}
H=\frac{1}{\sqrt{2}}
\bb
\phi^{+}\\
\left( v + h\cS{}  -s \sS{} \right) + i \phi^3\\
\eb\ , \quad
S=\left( f + h\sS{}  +s \cS{} \right)\ ,
\label{eqn:HSmix}
\end{align}
here, $v\sim246$ GeV is the Higgs vacuum expectation value, and $h$ and $s$ denote the physical Higgs ($m_h\simeq125$ GeV) and physical dilaton ($m_s\simeq 750$ GeV) fields, respectively. 
Mixing between $h$ and $s$ is parametrized by the angle $\theta_s$.

In addition to mixing between $H$ and $S$, there is also possible mixing between the top and its partner(s).\footnote{The term proportional to $\oli{T}_L u_{3R}$ can be 
rotated away and has not been included here. In an alternative basis, this term is kept and the $\oli{T}_R q_{3L} $ term is rotated away~\cite{Cacciapaglia:2011fx}.}
The analogous physical fields are denoted by $t$ and $t'$. The strength of this mixing is determined by an off-diagonal term in the $(t,T_i)$ mass matrix and is proportional 
to the coupling $y_i'$. For simplicity, we assume the mixing between the $N_T$ top partners $T_i$ and the top quark to be of equal magnitude and neglect
mixing among top partners themselves. In addition, we  consider the top partners to be nearly degenerate.\footnote{This assumption does not have any significant 
bearing on the production and decay of the Higgs or dilaton and therefore is of minimal consequence to the results in this paper.} Then, 
the general $(N_T+ 1)\times(N_T+ 1)$ mass matrix can be written as follows.

\begin{align}
		\bb
		\oli{u_{3L}} & \oli{T_{1 L}}\ ... & \oli{T_{N_T L}}
		\eb
		\bb
		m	& m' & ... &	m'\\
		0	& M  & ...  & 0\\
		\vdots&  & \ddots &\vdots \\
		0   &... &... & M 
		\eb
		\bb
		u_{3R} \\ T_{1R} \\\vdots \\ T_{N_T R}
		\eb.
		\label{eqn:mass-mat}
	\end{align}	
The assumption made on the form of the mass matrix here is reasonable since it does not strongly affect the phenomenology of dilaton production and decay, which is the main focus of our work here. Note that the top partner masses are nearly degenerate for this mass matrix.
As shown in Appendix~\ref{app:massmat},
all but one of the top partners are degenerate and have a mass $m_{t_i'}=M$ , while one of the top partners has a mass $m_{t_1'}\approx M + N_T m'$. However, a more complicated mass matrix could yield more complex relations between the masses of the top and its partners, which is beyond the scope of this work.

The renormalizable, linear scalar potential of the model has been given in Ref.~\cite{Abe:2012eu}, and we reproduce it below:
\begin{equation}
V(S,H)= \frac{m_S^2}{2} S^2 + \frac{\lambda_S}{4}S^4 + \frac{\kappa}{2}S^2 |H|^2 + m_{H}^2 |H|^2 + \frac{\lambda_H}{4}|H|^4\ .
\end{equation}
Note that the scalar potential has a $Z_2$ symmetry ($S\to -S$). This symmetry also holds for the physical fields when $\sS{}=0$, i.e. $s\to -s$. However, coupling of the dilaton to $T_i$ breaks this invariance.

From the discussion above, we see that in addition to SM parameters, the model introduces seven additional parameters; namely $M$, $f$, $y'$, $\lambda_S$, $\kappa$, 
$m_S$ and $N_T$. It is more useful to recast the parameters in terms of physical masses and mixing angles. For easy reference we list them below:
\begin{itemize}
	\item $m_s$: mass of the dilaton field, which we set to be $=750$ GeV.	
	\item $m_h$: mass of the physical Higgs field, which we set to be $=125$ GeV.
	\item $m_t$: mass of the top quark, which is set to $173$ GeV in this study.
	\item $m_{t'}$: mass of the top partner(s). Direct experimental limits set the value of $m_{t'}\gtrsim 780$ GeV~\cite{Chatrchyan:2013uxa,Aad:2015kqa}. 
Here, we set $m_{t_i'} = 1$ TeV. Note that changing $m_{t_i'}$ does not have a significant effect on Higgs ($h$) and dilaton ($s$) production cross sections and branching ratios.\footnote{Since we assume $2 m_{t'}$ is much larger than $m_h$ or $m_s$, the $h$ and $s$ cross section and branching ratios do not vary strongly with $m_{t'}$.}
	\item $\sS{}$: sine of the mixing angle $\theta_S$, which parametrizes the mixing between the Higgs and the dilaton fields, and is defined in Eq.~(\ref{eqn:HSmix}).

	\item $\sL{}$: sine of the mixing angle $\theta_L$, which parametrizes the mixing between $t$ and $t'$, defined for $N_T=1$ as follows:
		\al{
			\bb
			u_{3L}\\
			T_L
			\eb
			&=	\bb
			\cos\theta_L	&	\sin\theta_L\\
			-\sin\theta_L	&	\cos\theta_L
			\eb
				\bb
				t_L\\
				t'_L
				\eb.
		}
		When $N_T >1$, given the form of the mass matrix in Eq.~(\ref{eqn:mass-mat}) and assuming $m' \ll M$, 
		it is possible to characterize mixing between the top quark and its partner(s) by a single mixing angle ($\sL{}$) and by a $N_T\times N_T$ matrix, cf. Appendix~\ref{app:massmat}.
	\item $\eta=\frac{v}{f} N_T$: ratio of the two vacuum expectation values, multiplied by $N_T$.
	\item $N_T$: total number of additional vector like fermions, {\it i.e.} top partners  $T_i$.
\end{itemize}

We make further simplifications by only considering the limit $m_{t'}^2 \gg m_{t}^2$ and $m_{t'}^2 \gg m_{t}^2 \tan^2\theta_L$ . Note that the second condition ensures 
small mixing between the top and its partner(s).\footnote{Large mixing is constrained by oblique parameters~\cite{Abe:2012eu}. We will reanalyze these constraints for the general case, when $N_T \ge 1$.}
In this limit, we can write down the relation between the Lagrangian parameters and the physical parameters as follows:
\begin{align}
M\simeq m_{t'}\cL{} \ ,\\
y_{t} \simeq \frac{\sqrt{2}}{v}\frac{m_t}{\cL{}}\ ,\\
y'\simeq \frac{\sqrt{2}}{v}m_{t'}\sL{}\ .
\label{eqn:rel1}
\end{align}

Finally, in the large $m_{t'}$ and small mixing limit, the terms $ (m_t^2 \tan^2\theta_L/ m_{t'}^2) \to 0$ and one can simplify the coupling of the $h$ and $s$ fields as follows~\cite{Cao:2013cfa, Abe:2012eu}
\begin{eqnarray}
\tilde{C}_{hVV}&=&\frac{2 M_V^2}{v}\cS{}=C_{hVV}\frac{2 M_V^2}{v}\ ,\quad
\tilde{C}_{hff} =\frac{m_f}{v}\cS{}=C_{hff}\frac{m_f}{v}\ , \nn
\tilde{C}_{htt}&=&\frac{m_t}{v}(\cS{}\cL{2} + \eta\  \sL{2}\sS{})=\frac{m_t}{v}C_{htt}\ , \nn
\tilde{C}_{ht't'}&=&\frac{m_{t'}}{v}(\cS{}\sL{2} + \eta \cL{2}\sS{})=\frac{m_{t'}}{v}C_{ht't'}\ .
\label{eqn:h-coup}
\end{eqnarray}
Here, $V=\{W^{\pm},Z\}$ corresponds to massive Gauge bosons, and $ff$ to all fermions except the $t$ and $t'$.
Similarly for the dilaton field:
\begin{eqnarray}
\tilde{C}_{sVV}&=&\frac{2 M_V^2}{v}\sS{} =\frac{2 M_V^2}{v}C_{sVV}\ , \quad
\tilde{C}_{sff} =\frac{m_f}{v}\sS{}=\frac{m_f}{v} C_{sff}\ , \nn
\tilde{C}_{stt}&=&\frac{m_t}{v}(-\sS{}\cL{2} + \eta \sL{2}\cS{})=\frac{m_t}{v}C_{stt}\ ,\nn
\tilde{C}_{st't'}&=&\frac{m_{t'}}{v}(-\sS{}\sL{2} + \eta \cL{2}\cS{})=\frac{m_{t'}}{v}C_{st't'}\ .
\label{eqn:s-coup}
\end{eqnarray}
Note that the case of both $\sL{}=0$ and $\sS{}=0$ corresponds to the scenario when neither is there mixing between $h$ and $s$, nor between $t$ and $t'$. In this scenario $h$ has properties identical to the SM Higgs boson.
In the following section, we will present constraints on the model.
Note that for $N_T > 1$, the couplings $C_{htt}$, $C_{ht't'}$, $C_{st't'}$ and $C_{st't'}$ have additional powers of $\sL{}$ and $\cL{}$, cf. Appendix~\ref{app:massmat}.

\section{Model Constraints}
In this section, we determine constraints on the model from precision electroweak tests, Higgs signal strengths, single top production rate measurement and 
the $750$ GeV resonance production rates. 
\subsection{Constraints from Electroweak Precision Tests ($S$, $T$, $U$ parameters) }

\begin{figure}
	\centering
	\begin{minipage}{0.45\textwidth}
		\includegraphics[scale=0.6]{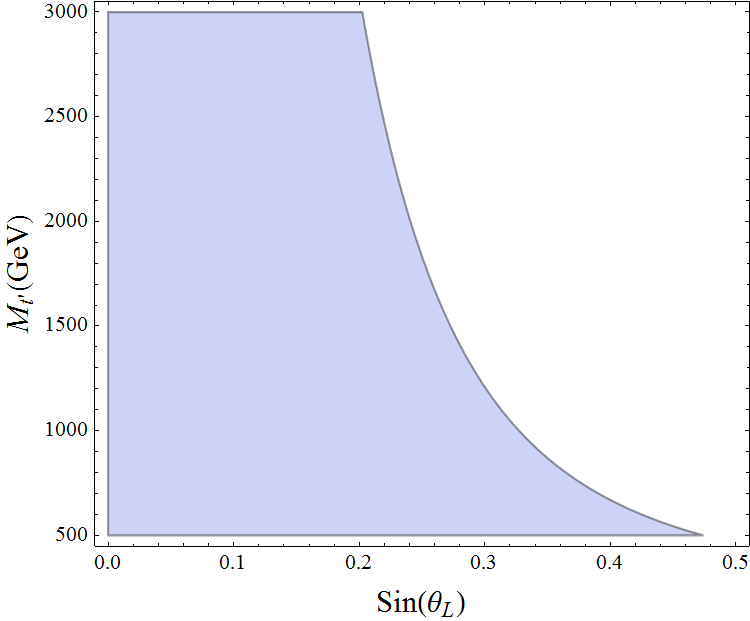}
		\caption{
			 $95\%$ CL limits from the $S$ and $T$ parameter in the $(m_{t'}-\sL{})$ plane, and for $\sS{}=0$ and $N_T=1$.
			 Shaded areas correspond to allowed regions of parameter space.
			\label{fig:STU-cL}
		}
	\end{minipage}
	\hfill
	\begin{minipage}{0.45\textwidth}
		\includegraphics[scale=0.6]{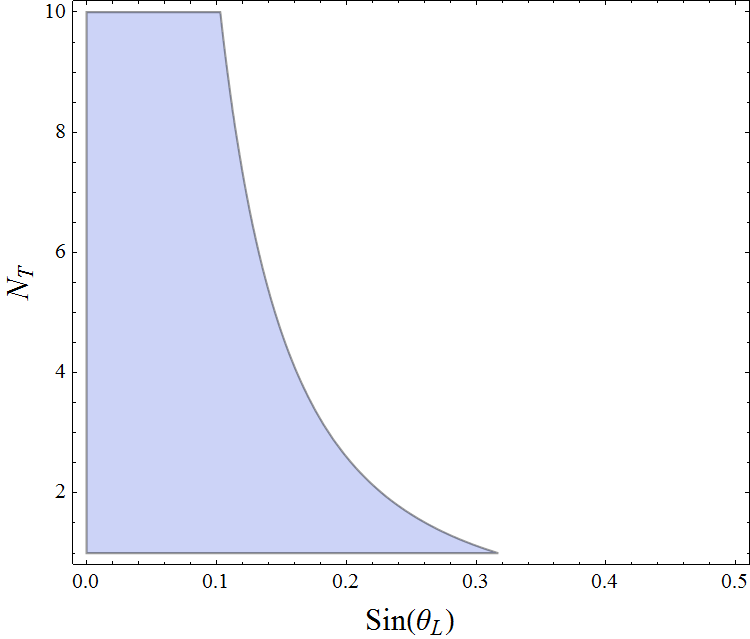}
		\caption{
			 $95\%$ CL limits from the $S$ and $T$ parameter in the $(N_T-\sL{})$ plane, and for $\sS{}=0$ and $m_{t'}=1$ TeV.
			 Shaded areas correspond to allowed regions of parameter space.
			\label{fig:STU-NT}
		}
	\end{minipage}
\end{figure}

The $S$, $T$ and $U$ parameters~\cite{Peskin:1990zt} of the model have been evaluated in Ref.~\cite{Abe:2012eu}. There are two contributions that need to be considered here, namely the contribution 
from mixing between the top and its partner(s), as well as from the scalar dilaton. This test strongly constrains mixing between the top and its partner(s). In Fig.~\ref{fig:STU-cL}, 
we show $95 \%$ Confidence Level (CL) upper bounds on the value of $\sL{}$ with varying $m_{t'}$, when $N_T=1$. 
In order to evaluate these bounds, we define a chi-squared function, $\chi^2_{STU}$, using
\begin{equation}
S= 0.05 \pm 0.09, \ \  T = 0.08 \pm 0.07 ,\ \  \rho_{ST}=0.91\ .
\end{equation}
Here, $\rho_{ST}$ is the correlation between $S$ and $T$, and the values of $S$ and $T$ are determined by setting $U=0$. We also set $\sS{}=0$; we will see below that the choice $\sin \theta_S = 0$ is best suited to explain the 750 GeV diphoton resonance
rate in this model.
When $m_{t'}=1$ TeV, we see that mixing between the top and its partner(s) is constrained, by the Electroweak precision tests, to the region where $\sL{} < 0.3$. We also observe that increasing $m_{t'}$ reduces the allowed magnitude of mixing between the top and its partner(s).

It is instructive to look at constraints on mixing between the top and its partners for the more general case when $N_T>1$. To this end, we calculate the $N_T$ dependence of the $T$ parameter and verify our results using {\tt package~X}~\cite{Patel:2015tea}. In general, each top partner can mix by different amounts. However, for simplicity, we only consider the special case of Eq.~(\ref{eqn:mass-mat}). As shown in Appendix~\ref{app:massmat}, this amounts to nearly degenerate top partners, cf. Eq.~(\ref{eq:app-mass-eval}), when the mixing between top quark and top partners is not too large. In Fig.~\ref{fig:STU-NT},
we show the $95 \%$ CL upper bounds on the value of $\sL{}$ with varying $N_T$ for  $m_{t'}=1$ TeV and $\sS{}=0$. We observe that the Electroweak precision tests constrain $\sL{}$ more strongly as $N_T$ increases.

\subsection{Constraints from Higgs signal strengths}
\label{sec:higgs-fit}

\begin{figure}[ht]
	\centering
	\begin{minipage}{0.45\textwidth}
	\includegraphics[scale=0.6]{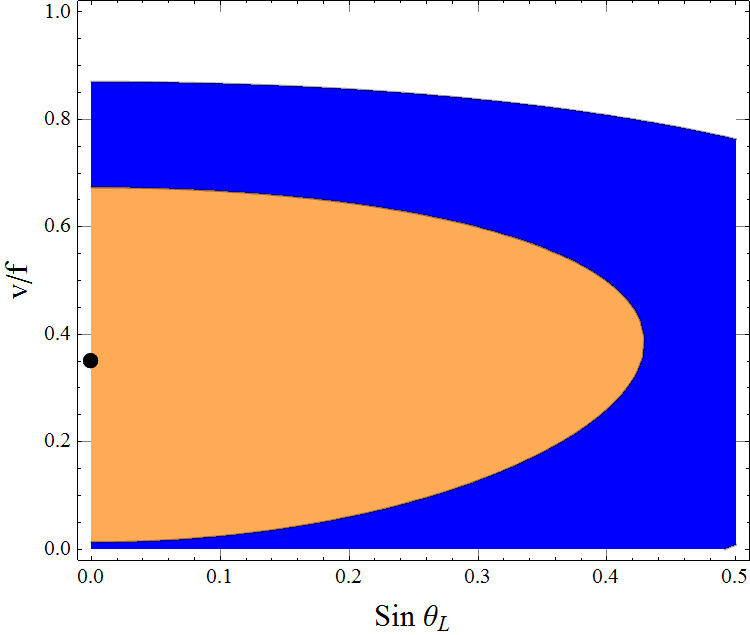}
	\caption{
		 Contours of $1\sigma$ and $2\sigma$ deviation of $\Delta\chi^2$ in the $(\sL{} - \frac{v}{f})$ plane, constrained by Higgs signal strength data only. Here $(\sS{}=0.31,\sL{}=0,\frac{v}{f}=0.35)$ is the best fit value and is denoted by a black dot. 
	}
	\end{minipage}
	\hfill
	\begin{minipage}{0.45\textwidth}
		\includegraphics[scale=0.6]{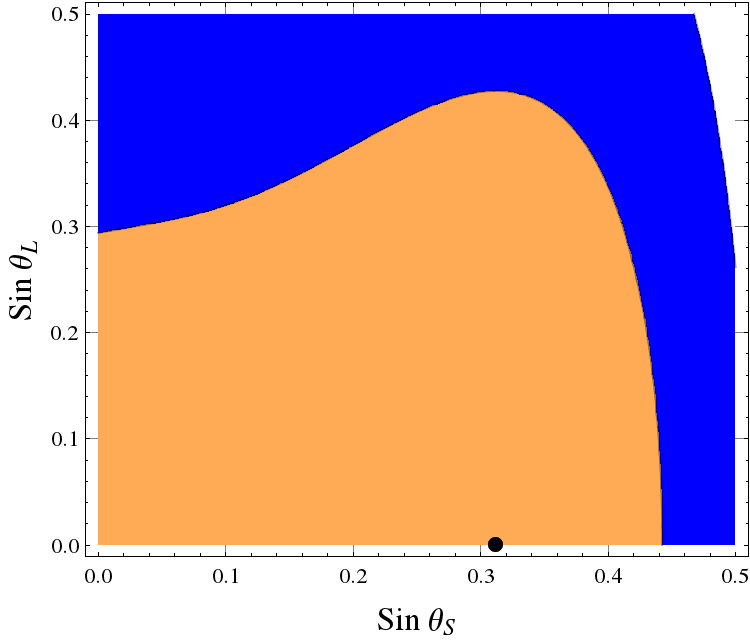}
		\caption{
		$1\sigma$ and $2\sigma$ contours of $\Delta\chi^2$ in the $(\sS{} - \sL{})$ plane, constrained by Higgs signal strength data only. Here $(\sS{}=0.31,\sL{}=0,\frac{v}{f}=0.35)$ is the best fit value and is denoted by a black dot.
	\label{fig:chi2-cont}}
    \end{minipage}
\end{figure}

Higgs signal strengths have been measured using data from the $7$ and $8$ TeV runs of LHC. These signal strengths define properties of the observed 125 GeV Higgs boson. Examining these signal strengths provides insights into the amount of mixing allowed between the Higgs boson and the 750 GeV dilaton. Therefore, in this section, we perform fits to MDM paramters using Higgs data.

The signal strengths reported by ATLAS and CMS are defined as
\begin{equation}
\hat{\mu}_i = \frac{n_{\rm exp}^i}{n_{\rm SM}^i}~,
\label{eq:mui_exp}
\end{equation}
where $n_{\rm exp}^i$ is the number of events observed in the channel $i$ and $n_{\rm SM}^i$
is the expected number of events as predicted in the SM. In order to compare the MDM predictions
with the experimentally derived $\hat \mu_i$, we define (as usual)
\begin{equation}
\mu_i = \frac{n_{\rm th}^i}{n_{\rm SM}^i} = 
\frac{\Sigma_p \sigma_p^{MDM} \epsilon_p^i}{\Sigma_p \sigma_p^{\rm SM} \epsilon_p^i} 
\times \frac{\mathcal B_i^{MDM}}{\mathcal B_i^{\rm SM}}~.
\label{eq:mui}
\end{equation}
Here $\sigma_p$ is the production cross section of the $125$ GeV Higgs boson in the $p^{th}$ channel and $\mathcal B_i$ corresponds to its branching ratio in the $i^{th}$ channel. $\epsilon_p^i$ correspond to the fraction that each production channel contributes in the search for the Higgs in its $i^{th}$ decay channel.
Fits to $\mu_i$ are performed by minimizing the $\chi^2$ function defined as 
\begin{equation}
\label{eq:chi2}
\chi^2 =\sum\limits_{i}\left(\frac{\mu_i -\hat{\mu}_i}{\hat \sigma_i}\right)^2.
\end{equation}

After setting $m_{t'}=1$ TeV and $N_T=1$, we are left effectively with three model parameters; $\sS{},\ \sL{}$ and $\frac{v}{f}$. Note that varying  $m_{t'}$ has a minimal effect on the fits since 
$m_{t'}^2 \gg m_h^2$ and therefore the contribution of $t'$ to Higgs decays can be calculated in the heavy quark limit to a good approximation. Using 
Higgs data, listed in the Appendix of Ref.~\cite{Boudjema:2015nda}, we perform a fit over the three model parameters $(\sS{},\sL{},\frac{v}{f})$ and find the best fit to be at $(0.31,0.0,0.35)$ 
with the $\chi^2/d.o.f = 27.6/24$. In Fig.~\ref{fig:chi2-cont},  $1\sigma$ (one standard deviation) and $2\sigma$ contours of $\Delta\chi^2$ are shown. The left plot shows the contours in the $(\sL{} -\frac{v}{f})$ plane, 
with $\sS{}=0.31$, whereas the right plot shows the same in the $(\sS{}-\sL{})$ plane, with $\frac{v}{f}=0.35$. The black dots indicate the best fit points on the plots.
Both plots highlight the preference for small or no mixing: i.e., $\sin\theta_L \sim 0$ 
and $\sin\theta_S \sim 0$.

\begin{figure}
	\centering
	\includegraphics[scale=0.6]{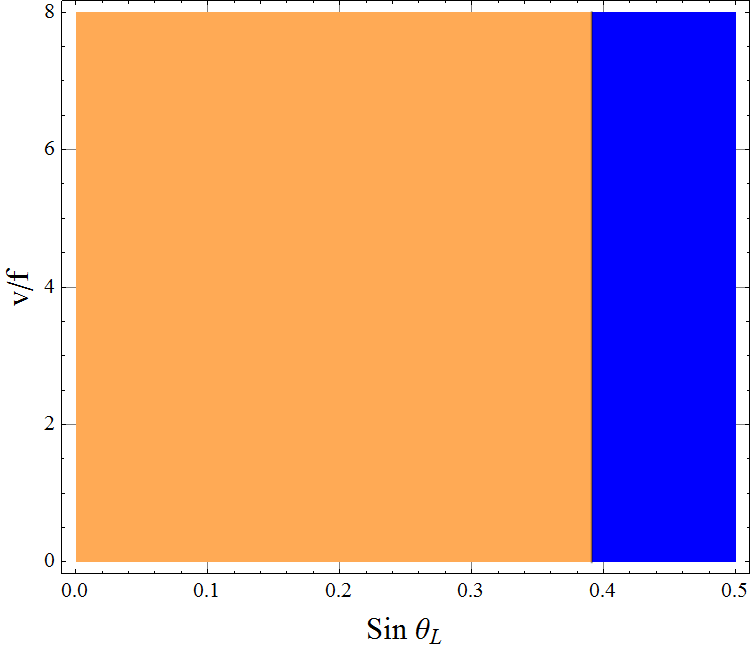}
	\caption{ $1\sigma$ and $2\sigma$ contours of $\Delta\chi^2$ in the $(\sL{} - \frac{v}{f})$ plane. Here the best fit occurs at the line of $\sL{}=0$, with $\sS{}=0$ (see text for details). Fits are performed using Higgs data alone.
\label{fig:chi2-cont2} }
\end{figure}

In fact, we will see in the next section that the choice $\sS{}=0$, is best suited to explain the diphoton resonance rates. Keeping this in mind, we fix $\sS{}=0$ and perform fits to the Higgs signal strength data by allowing $\frac{v}{f}$ and $\sL{}$ to vary freely. We find that the $\chi^2$ at the best fit value is $\chi^2/d.o.f = 28.5/25$. In Fig.~\ref{fig:chi2-cont2}, we 
show $1\sigma$ and  $2\sigma$ contours of the fit in the $(\sL{}-\frac{v}{f})$ plane. $\sL{}=0$ corresponds to the best fit and $\frac{v}{f}$ is not constrained by the fit. When $\sS{}=0$, the $\frac{v}{f}$ dependence of Higgs signal strengths drops out, resulting
in a flat behavior of the $\chi^2$ function in the $\frac{v}{f}$ direction. This results in the vertical contours seen in Fig.~\ref{fig:chi2-cont2}. 

Constraints on $\sL{}$ appear to be fairly weak, with all values of $\sL{}$ being allowed within $3\sigma$. This behavior can be understood from the following
argument. The $\sL{}$ dependence of the $\chi^2$ function arises through three sources: decay of Higgs to a pair of gluons (as well as gluon fusion production), 
decay of Higgs to a pair of photons, and $t\bar{t} h$ production. For the first two sources, the $\sL{}$ dependence is weak, since effectively the amplitudes for these 
decays are $\cL{2}A_{t} + \sL{2}A_{t'}$. Here $A_{f}$ are loop functions defined in the Appendix~\ref{app:hdecay}. Since both $m_{t}$ and $m_{t'}$ are larger than $m_h$, $A_{t}\sim A_{t'}$ and the $\sL{}$ 
dependence nearly vanishes from $h\to gg$ and $h \to \gamma \gamma$ processes. This leaves only $t\bar{t}h$ production with a $\sL{}$ dependence. Since $t\bar{t}h$ production has been measured with large errors, therefore the constraint on $\sin\theta_L$ from the current measurement is relatively weak. As discussed earlier, a stronger constraint on $\sL{}$ has been imposed by the Electroweak precision tests, cf.~Figs.~\ref{fig:STU-cL} and \ref{fig:STU-NT}. 

\subsection{Constraints from Single Top Production}

Top quarks are copiously produced at the LHC. An important observable that would be modified by this model is the single top cross section. The current
constraints from ATLAS and CMS already put a strong constraint on the product of the mixing between the top quark and top partner(s) and $|V_{tb}|$. The current most precise measurements
from ATLAS and CMS yield $|V_{tb}| > 0.88$ and $|V_{tb}| > 0.92$, respectively, at the $95\%$ confidence level (CL) ~\cite{1510.03764,1403.7366}. The single top production cross section in this new model is proportional to the product
$(|V_{tb}| \cL{})^2$.

\begin{figure}[ht]
 \centering
 \begin{minipage}{0.45\textwidth}
  \includegraphics[scale=0.5]{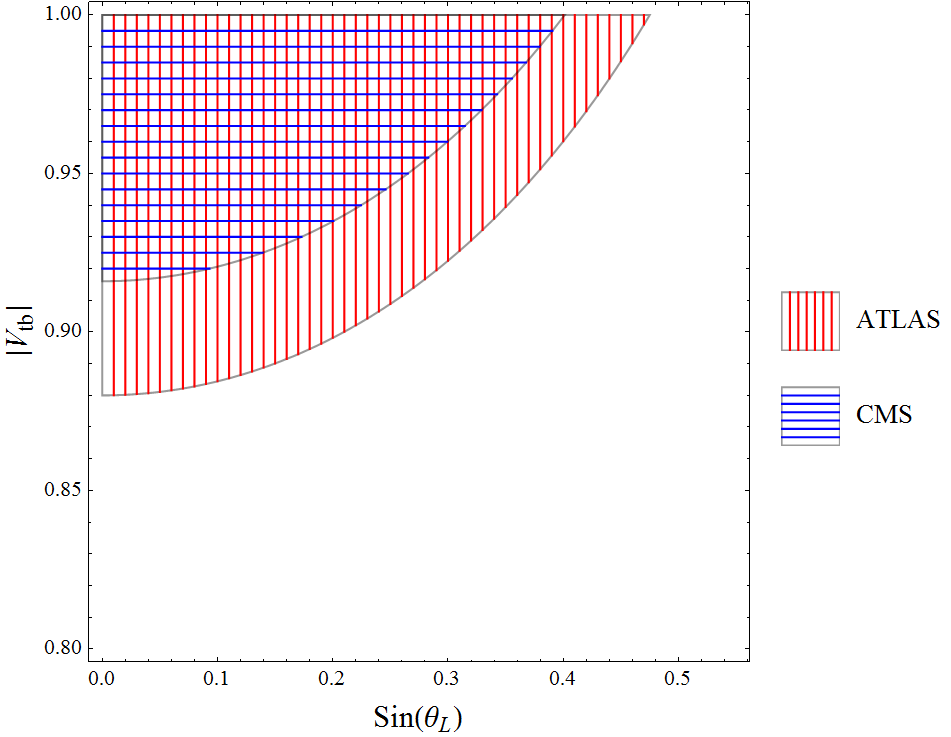}
  \caption{Allowed (shaded) region from ATLAS (darker shaded) and CMS (lighter shaded) measurement of single top production, independent of $m_{t'}$,  for $N_T=1$. The CMS constraint overlaps the ATLAS constraint.}
  \label{fig:s-top-nt1}
 \end{minipage}
 \hfill
 \begin{minipage}{0.45\textwidth}

 \includegraphics[scale=0.5]{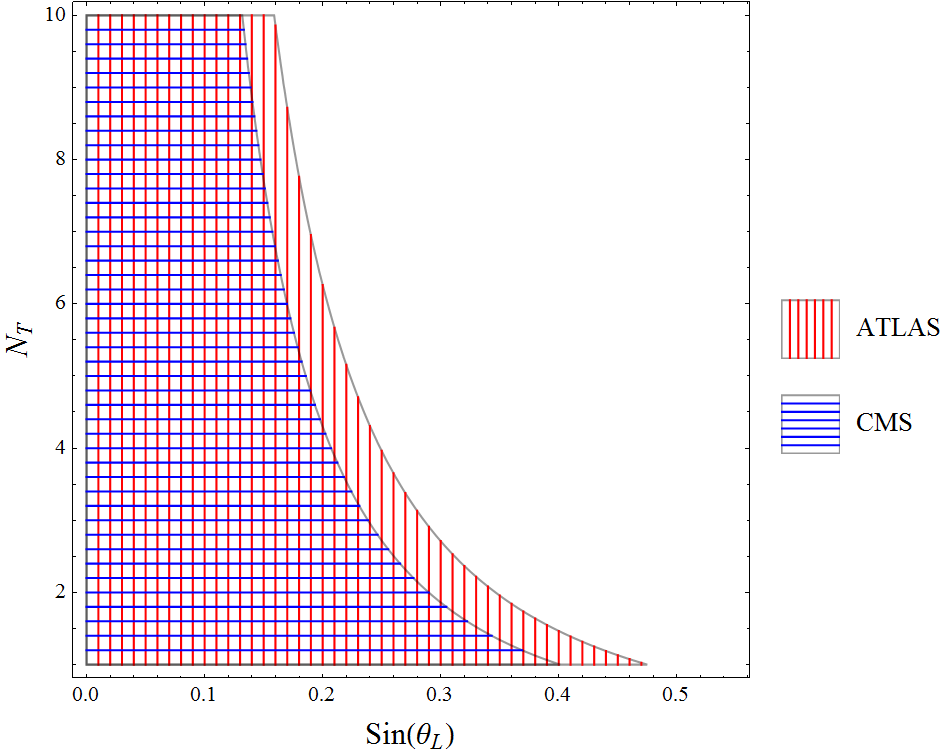}
 \caption{Allowed (shaded) region from ATLAS (darker shaded) and CMS (lighter shaded) measurement of single top production, independent of $m_{t'}$,  assuming $|V_{tb}|$=1. The CMS constraint overlaps the ATLAS constraint.}
 \label{fig:s-top-ntN}
 \end{minipage}
\end{figure}

In Fig.~\ref{fig:s-top-nt1}, we show $95\%$ CL allowed regions in the $(|V_{tb}| - \sL{})$ plane, when $N_T=1$. We observe that the constraints on $\sin(\theta_L)$ are comparable to those from the $T$ parameter constraints, as seen in Figs.~\ref{fig:STU-cL} and \ref{fig:STU-NT}.
In Fig.~\ref{fig:s-top-ntN} we show constraints from single top production in the $(N_{T} -\sL{})$ plane, after setting $|V_{tb}|=1$. It is evident that allowed regions of $\sL{}$ reduce as $N_T$ increase.
Currently, constraints from the $T$ parameter are only slightly stronger than
that from the single top production. However, with the high luminosity runs of the HL-LHC $(3 {\rm ab}^{-1})$, it is
expected that there will be 15 million single top events. The estimated uncertainty on single top production should be reduced to $3.8\%$ by the end of the LHC
run~\cite{Schoenrock:2013jka}. This would imply a reduction on the uncertainty of approximately a factor of $2 \sim 3$ in $(|V_{tb}| \cos\theta_L)^2$. If we choose $|V_{tb}|=1$,
then the limits would be $\sL{}\lesssim 0.2$ (which is stronger than the Electroweak precision test constraints) at the end of the running of HL-LHC, ignoring changes in theortical and systematic uncertainties. Moreover, constraints from the single top measurement is independent of the top partner mass.

\subsection{Dilaton production and Decay}
Production of the dilaton of mass $m_s =750$ GeV proceeds through a loop induced process of gluon fusion. All other production modes are expected to be 
sub-dominant, especially in the small mixing $(\sS{}\sim 0)$ scenario.
We estimate the cross section for production and decay of the dilaton using the Narrow Width Approximation (NWA). Since we set $\sS{}=0$, the dilaton decays primarily through loop induced processes, and therefore we find that the ratio of its decay width to mass is $\Gamma_s/m_s \sim 10^{-4}$.\footnote{The decay width of the of the dilaton is proportional to $N_T^2/f^2$.} The NWA is thus a good approximation for production of the dilaton. 

The production cross section in the NWA can be written down as follows.
\begin{equation}
\sigma(gg \to s \to \gamma \gamma) = 
16\pi^2 \cdot {\cal N} \cdot \frac{ \Gamma_s}{m_s} \cdot
BR(s\to \gamma \gamma) \cdot BR(R\to gg) \cdot \left[ \frac{d L^{gg}}{d\hat{s}}\right]_{\hat{s} = m^2_s}~,
\label{eq:simplest}
\end{equation} 
where ${\cal N}$ is a ratio of spin and color counting factors
\begin{equation}
{\cal N} = \frac{N_{s}}{N_{g} N_{g}} \cdot	
\frac{C_{s}}{C_g C_g},
\end{equation}
where $N$ and $C$ count the number of spin- and color-states, respectively, for initial state gluons and the resonant dilaton, with $N_s=1,N_g=2,C_g=8$ and $C_s=1$. In order to 
compute the total cross section at $13$ TeV LHC,
we multiply the above mentioned partonic cross section with the gluon luminosity $L_{gg}$. We evaluate $L_{gg}$ using the {\tt CT14LO} parton distribution
function~\cite{Dulat:2015mca} and the {\tt LHAPDF} package~\cite{Buckley:2014ana}. We also evaluate the NNLO $K$-factor using the {\tt SuSHi} program~\cite{Harlander:2012pb} in the infinite quark mass limit with the
{\tt CT14NNLO} PDF central set~\cite{Dulat:2015mca}. We set  the renormalization and factorization scales to be $\mu_R=\mu_F=750$ GeV. We find the $K$-factors to be 
$K_{NNLO/LO}^{13 TeV}\sim 2.6$ and $K_{NNLO/LO}^{8 TeV}\sim 2.7$, respectively, at the $13$ TeV and $8$ TeV LHC, which will be included in the following analysis of dilaton production rates.
An interesting feature of this model is that when all mixings are set to zero (i.e. $\sL{}=0$ and $\sS{}=0$), all decays are loop induced. The branching ratios are fixed by the $SU(3)_c$ and $U(1)_Y$ charges of the the top partners. The decay width on the other hand is proportional to $N_T^2/f^2$. From the equation above we see that in order to increase the cross section one could either increase the decay width by increasing $N_T$ or decreasing $f$. However, we will see later that small values of $f$ will lead to large Yukawa couplings of the top partners and therefore to issues of perturbativity.
\begin{figure}
	\centering
\begin{minipage}{0.45\textwidth}
	\includegraphics[scale=0.45]{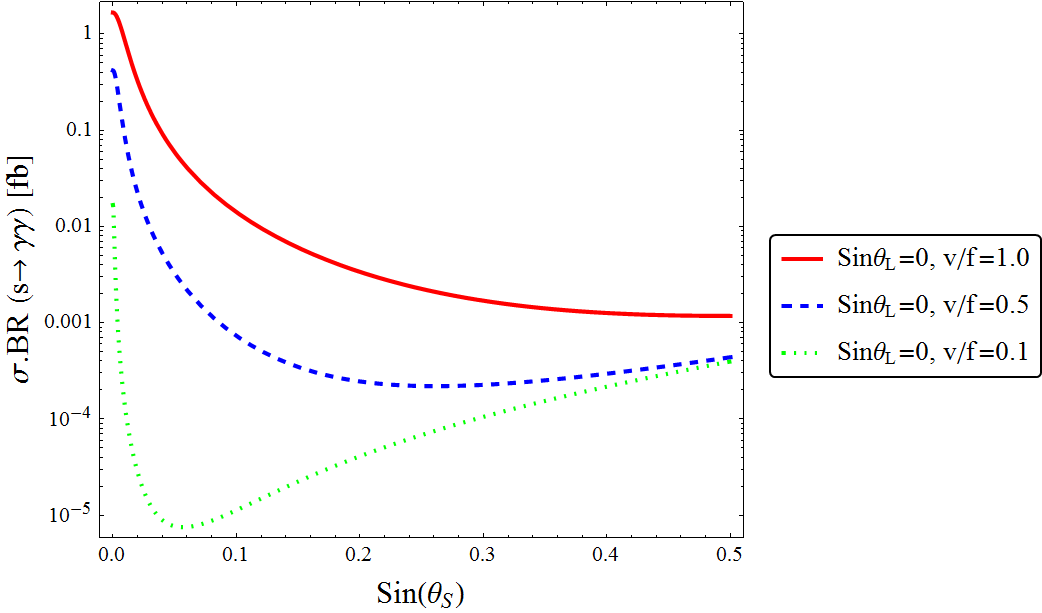}
	\caption{Variation of the next-to-next-to leading order production cross section times branching ratio $(\sigma(pp \to s)\cdot \text{BR}(s\to\gamma\gamma))$ of the dilaton ($m_s=750$ GeV) for $13$ TeV LHC, with $\sL{}=0$.
	Here we set $N_T=1$. \label{fig:csBR}}
\end{minipage}
\hfill
\begin{minipage}{0.45\textwidth}
	\includegraphics[scale=0.45]{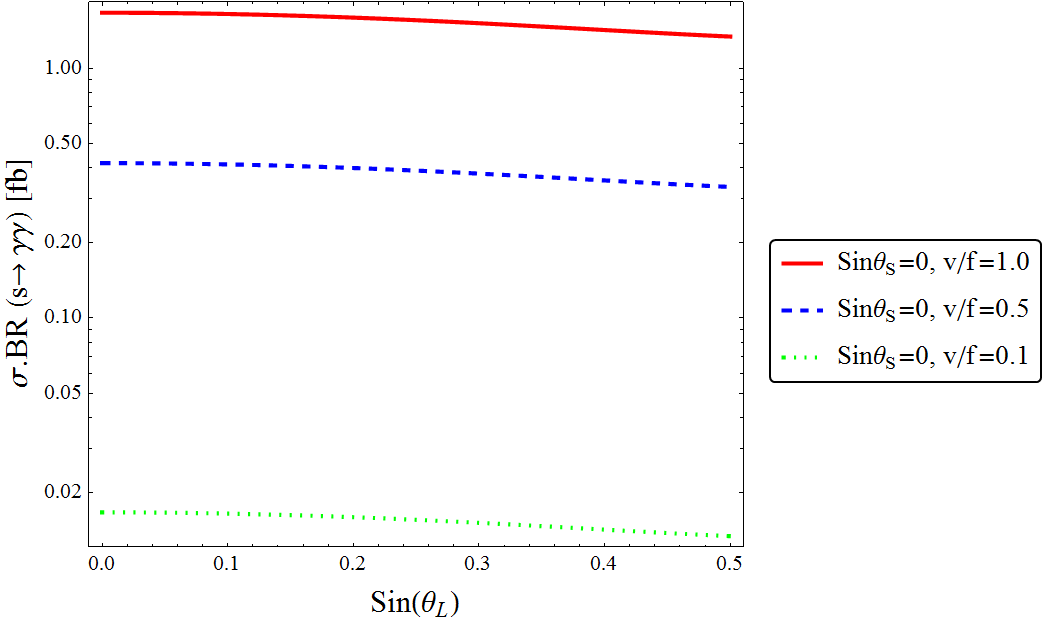}
	\caption{Variation of the next-to-next-to leading order production cross section times branching ratio $(\sigma(pp\to s)\cdot \text{BR}(s\to\gamma\gamma))$ of the dilaton ($m_s=750$ GeV) for $13$ TeV LHC, with $\sS{}=0$. Here we set $N_T=1$.} \label{fig:csBR2}
\end{minipage}
\end{figure}

The expressions for the decay width of the dilaton are given in
Appendix~\ref{app:sdecay}. In Fig.~\ref{fig:csBR}, we show the variation of $\sigma(pp \to s \to \gamma \gamma)$ with the 
Higgs-dilaton mixing angle $\sS{}$ for three different values of $\frac{v}{f}=\{1,0.5,0.1\}$ and $\sL{}=0$. When $\frac{v}{f}=1$, we find, in the limit of
$\sS{}\to 0$, the NNLO cross section $\sigma(pp \to s \to \gamma \gamma) \to 1.1 $ fb. As mixing between $h$ and $s$ becomes non-zero ($\sS{}\ne 0$), there is  a dramatic fall in the $\sigma$, as can be seen in Fig.~\ref{fig:csBR}. This behavior is due to the large negative
interference arising from the $W$-boson loop contribution to the diphoton decays of the dilaton. Furthermore, with non-zero values of $\sS{}$, tree level decays of $s\to VV$ where $V=\{W^{\pm},Z\}$ are allowed and become the dominant decay modes, further reducing the $s\to \gamma \gamma$ branching ratio, cf.~Fig.~\ref{fig:app:BR-s} in Appendix~\ref{app:bratio}.

In the rest of this paper we therefore restrict ourselves to the scenario where $\sS{}=0$. We are therefore left with three parameters in the model: $\sL{},f$ and $N_T$. When we set $\sS{}=\sL{}=0$, the dilaton $s$ can decay only through four different loop processes $s\to \gamma\gamma$, $s\to Z\gamma$, $s\to ZZ$ and $s\to gg$. 
When $\sL{}\ne 0$, then the tree level $s\to t \bar{t}$  decay channel also becomes available, though suppressed by powers of $\sL{4}$ in the small mixing angle limit.\footnote{Mixing with the top quark also allows for the decay channel $s \to W^+ W^-$ through a triangle diagram. However, since the mixing angle $\sL{}$ 
is strongly constrained by electroweak precision tests to a small value, we may safely neglect this decay channel.} 
Consequently, the total decay width of the dilaton is expected to be very small.
Branching ratios to these various channels are shown in Appendix~\ref{app:bratio}.

In Fig.~\ref{fig:csBR2}, the variation of $\sigma(pp \to s \to \gamma \gamma)$ with the $t-t'$ mixing angle $\sL{}$ is shown. We see that there is only a weak dependence on the mixing angle $\sL{}$. This behavior, as explained earlier, is due to the fact that for the choice of $\sS{}=0$, 
the $\sL{}$ dependence of the cross section almost vanishes.  As expected, one can see that the cross section simply depends quadratically on $\eta=N_T v/f$.

In order to constrain the model, we use the following values of cross sections
\begin{itemize}
\item $\sigma(pp \to s \to \gamma \gamma) = 6.26 \pm 3.23 $ fb \cite{1512.04939, ATLAS-CONF-2015-081, CMS-PAS-EXO-15-004}.
\item $\sigma({pp \to s \to Z \gamma }) < 8.2 $ fb \cite{Aad:2014fha}.
\item $\sigma({pp \to s \to g g }) < 2200 $ fb \cite{CMS-PAS-EXO-14-005}.
\item $\sigma({pp \to s \to Z Z }) < 19 $ fb \cite{Aad:2015kna}.
\item $\sigma({pp \to s \to t \bar{t}  }) < 700 $ fb \cite{Aad:2015fna}.
\end{itemize}

\begin{figure}
	\centering
	\begin{minipage}{0.45\textwidth}
		\includegraphics[scale=0.5]{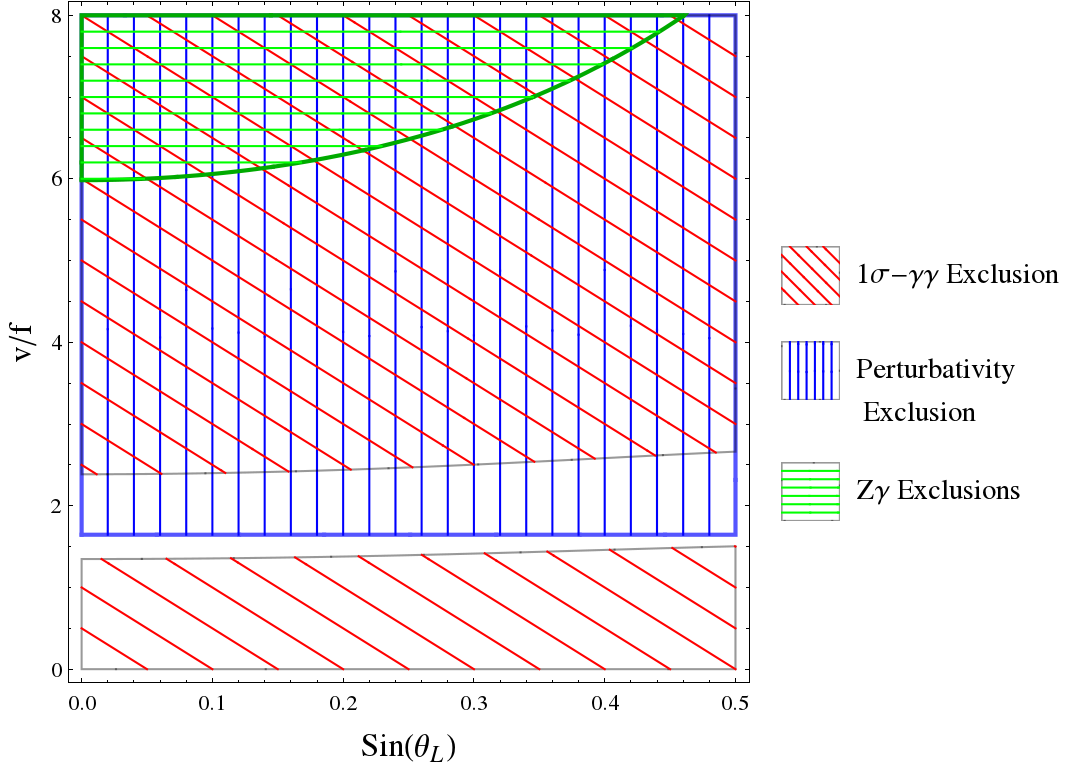}
		\caption{ Constraints from $pp \to s\to (\gamma \gamma,Z \gamma)$ production rates of the dilaton as well as perturbativity requirement($\lambda_S \le 4 \pi$). The shaded regions are excluded. 
			  Here $N_T=1$.  \label{fig:other-limits}}
	\end{minipage}
	\hfill
	\begin{minipage}{0.45\textwidth}
		\includegraphics[scale=0.5]{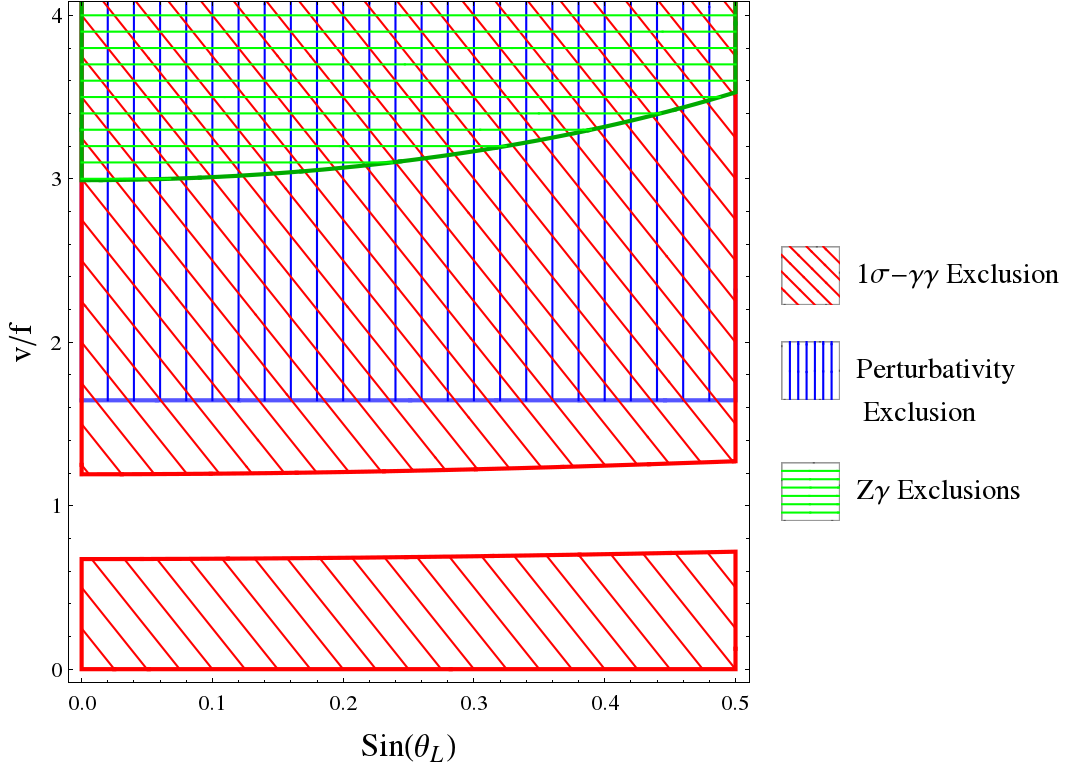}
		\caption{Constraints from $pp \to s\to (\gamma \gamma,Z \gamma)$ production rates of the dilaton as well as perturbativity requirement($\lambda_S \le 4 \pi$). The shaded regions are excluded.
			  Here $N_T=2$.  \label{fig:other-limits-NT2}}
	\end{minipage}
\end{figure}

In Fig.~\ref{fig:other-limits}, we set $N_T=1$ and show, in the $(\sL{}-\frac{v}{f})$ plane, constraints from one sigma exclusion limits from $\sigma(pp \to s \to \gamma \gamma)$ and $\sigma(pp \to s \to Z \gamma)$ and perturbativity requirement $(\lambda_S \le 4 \pi)$. Similarly, in Fig.~\ref{fig:other-limits-NT2}, we show the same exclusions for the case when $N_T=2$.
We do not show constraints from $s\to gg$, $s\to ZZ$ and $s\to t \bar{t}$ decay modes in these figures since they are very weak. The allowed region, that explain the diphoton excess to within one standard deviation, corresponds to the unshaded white regions in the figure.
In  Fig.~\ref{fig:other-limits}, where $N_T=1$, we see that large values of $\frac{v}{f}$ are necessary in order to explain the diphoton excess. However, such large values of $v/f$ (around $2$) are mostly ruled out by the requirement of perturbativity $(\lambda_S \le 4 \pi)$ and only a small region of parameter space remains to explain the signal. On the other hand, in  Fig.~\ref{fig:other-limits-NT2}, where $N_T=2$, we see that smaller values of $\frac{v}{f}$ (hence larger values of $f$) can explain the diphoton excess, thus evading bounds from perturbativity ($\lambda_S \le 4\pi$).

\begin{figure}
	\centering
		\includegraphics[scale=0.55]{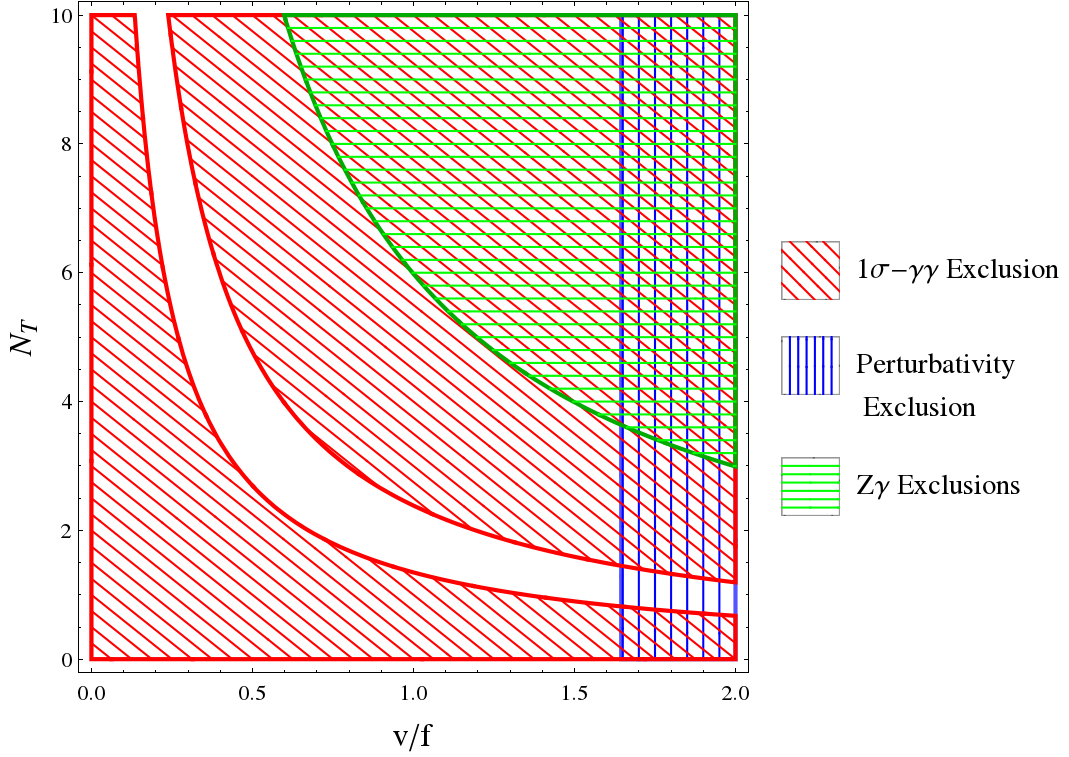}
		\caption{ Constraints from the $pp \to s\to (\gamma \gamma,Z \gamma)$ production of the dilaton as well as well as perturbativity of the of the dilaton's quartic coupling($\lambda_S \le 4 \pi$). 
		Here we set $\sL{}=\sS{}=0$.  \label{fig:NTvar1}}
\end{figure}

In Fig.~\ref{fig:NTvar1}, we show the same set of constraints in the $(\frac{v}{f} -N_T)$ plane while setting $\sL{}=0$. We see that the model is able to explain the diphoton excess and avoid other constraints when $N_T\ge 1$. Further, as $N_T$ increases, the value of $f$ required to explain the diphoton excess increases (denoted by the unshaded white region in the figure). 

As discussed in Appendix~\ref{app:vac-stab}, although the MDM, in its current form, can explain the diphoton excess, additional new dynamics is required in the TeV region to stabilize the vacuum. It is not necessary that the new dynamics will contribute to the diphoton signal. It is only required that it contributes positively to the running of the coupling $\beta_{\lambda_S}$. We therefore do not speculate the nature of this new dynamics as there are a vast number of possibilities.

It is interesting to observe that this model has very definite predictions for $s\to Z\gamma$, $s\to ZZ$ and, $s\to t\bar{t}$ branching ratios.\footnote{
See Appendix~\ref{app:sdecay} and \ref{app:bratio} for details. } 
If the diphoton excess is confirmed in future runs of the LHC , then a way to further test this model is to determine these branching ratios. We discuss this in further detail later in the paper.

\section{Phenomenology of MDM at LHC}
In this section we discuss the phenomenology relevant for future LHC runs in the context of the Minimal dilaton model with its constraints as derived in earlier sections. 
In light of the $750$ GeV excess, the constraints on MDM can be summarized as follows:
\begin{itemize}
	\item Mixing between the $125$ GeV Higgs boson and the dilaton $s$ is effectively nil, i.e. $\sS{}\sim0$. We will consider phenomenology of the the MDM strictly in the case when $\sS{}=0$.
	\item Precision constraints and single top measurements require the mixing between the top and its partner(s) to be small (i.e. $\sL{}\ll 1$).
	\item We find the Minimal Dilaton Model with $0.2\leq v/f\leq 1.0$ and $1< N_T<10$ is a valid effective theory up to the scale of $\sim m_{t'}$. Not only
              does it explain the diphoton signal, but also satisfies all the constraints from both experimental data and the requirement for perturbativity. For $N_T > 10$, the beta function of the strong coupling constant becomes positive such that the QCD interaction is no longer asymptotically free. Hence, we do not consider this case.  
	\item The top partner mass is constrained by requiring the perturbativity of the Yukawa coupling $y_T \sim  M/f < 4 \pi /\sqrt{N_c}$ GeV, where $N_c =3$ counts the number of color states of the top partner(s).
	Since the dilaton is entirely responsible for generating the mass of the top partner(s), $m_{t'}$ is bounded from above: $m_{t'} < (1800 f/v ) $ GeV, in the small mixing
	angle approximation. We therefore expect to find top partner(s) with masses in the range of $2-9$ TeV given the bounds from the previous point.
\end{itemize}

Given the approximate constraints summarized above, we proceed to discuss the phenomenology of the model.
\subsection{Determination of dilaton branching ratios}
If the diphoton resonance is confirmed at the future runs of the LHC, the next step is to determine the ratio of branching ratios of the dilaton into various decay channels.
Since we consider the limit $\sS{}=0$ and $\sL{}\ll 1$, the possible decay channels of the  dilaton are 
\begin{itemize}
	\item $s \to  g g$: In the limit $\sS{}=\sL{}=0$, $\Gamma(s \to  g g)/\Gamma(s \to  \gamma \gamma) \sim 9\alpha_s^2/ 8\alpha^2 \sim 14$. This decay channel can be investigated through dijet decays.\footnote{ That the dijets are indeed of gluonic origin may be investigated with the aid of tools such as jet energy profiles. See, for example, ~\cite{Chivukula:2014pma}. }
	\item $s \to Z \gamma$: In the limit $\sS{}=\sL{}=0$, $\Gamma(s \to  Z \gamma)/\Gamma(s \to  \gamma \gamma) \sim 2 \tan^2\theta_w \sim 0.6$.
    \item $s \to ZZ$: In the limit $\sS{}=\sL{}=0$, $\Gamma(s \to  Z Z)/\Gamma(s \to  \gamma \gamma) \sim  \tan^4\theta_w \sim 0.1$.
	\item $s\to t \bar{t}$: This process is proportional to $\sL{4}$ and is therefore quite small. For $\sL{}=0.3$, $\Gamma(s \to  t \bar{t})/\Gamma(s \to  \gamma \gamma) \sim 0.06$.
	\item $s \to W^+ W^-$: This decay channel has a non-zero partial decay width only when $\sL{}\ne 0$. Further, since it is proportional to $\sL{4}$, it is expected to be highly suppressed, as compared to the other decay modes, cf.~Appendix~\ref{app:sdecay}.
    \item $s \to hh$:  Similar to the $WW$ channel, this decay mode has a non-zero partial width only when $\sL{}\ne 0$  and is expected to be very small.
\end{itemize}
The decay width of the dilaton, as mentioned earlier is expected to be small. In fact, when $\sS{}=0$, the ratio of decay width to mass of the dilaton  ($\Gamma_s/m_s$) is of order $10^{-4}$. In this model, this ratio, can be large ($\sim 0.3$) when $\sS{}=1$. However, $\sS{}\ne 0$ is unfavorable for explaining the diphoton excess. The decay width also increases as $N_T^2$ when $\sS{}=0$, however, the range of variation is small.  For $N_T = 1$ we find $\Gamma / M \sim 6\times10 ^{-5}$ and for $N_T=10$ we find $\Gamma / M \sim 6\times10 ^{-3}$.

\subsection{Determination of top partner mass and branching ratios}
Electroweak decay modes of top partners are possible when mixing with top quarks is present ($\sL{}\ne0$).
In such a scenario, the $t'$ has four possible decay modes: $t' \to W t$, $t' \to Z t$, $t' \to h t$ and, if kinematically allowed, $t' \to s t$. 
The first three decay widths are related through the Goldstone Equivalence Theorem~\cite{Cornwall:1974km,Lee:1977eg,Gounaris:1986cr,Chanowitz:1985hj,Yao:1988aj} 
and one finds that $\Gamma(t' \to W t) \simeq 2 \Gamma(t' \to Z t) \simeq 2 \Gamma(t' \to h t) $~\cite{Dobrescu:2009vz}. For the fourth decay channel 
(assuming $m_{t'}\gg(m_{t},m_{s})$ and $\sL{}\ll 1$), we have $\Gamma(t' \to s t)\simeq (v/f)^2 \Gamma(t' \to h t) $. Therefore, the decay channel is suppressed not 
only through phase space factors but also through $(v/f)^2$. 
Hence, in the context of MDM, one could follow the traditional strategy of searching for top partners.\footnote{
See, for example, Ref.~\cite{Chatrchyan:2013uxa} and references therein.}

As explained earlier, one expects to find top partners with masses in the range of $\sim 2-9$ TeV. Top partners can be searched for at hadron colliders, either through pair or single production. For $14$ TeV LHC with $3000\, {\rm fb^{-1}}$ integrated luminosity, $5\sigma$ discovery is possible up to $m_{t'}\sim 1.5$ TeV, whereas a $33$ TeV machine with the same amount of integrated luminosity could discover a top partner with $m_{t'}\sim 2.4$ TeV~\cite{Agashe:2013hma}. For a 100 TeV machine it is possible to set limits as large as $m_{t'} \gtrsim 6$ TeV~\cite{Matsedonskyi:2014mna}.  

One interesting aspect pertaining to the particular form of the mass matrix used in this study is that for $N_T >1$, the top partner masses are not all degenerate. Furthermore, from earlier discussion we learned that in order to explain the diphoton signal, $N_T$ could be greater than one. Therefore, one might expect to find top partners of same charge, but different masses. If $m'$, cf.~Eq.~(\ref{eq:app-mass-eval}), is large, the mass difference between the heaviest top partner and the degenerate top partners may be large enough to open up decay channels of the form $t_1' \to t_2' + h/Z $. This presents itself as a novel signal albeit a weaker one in comparison to decay modes discussed earlier.
\subsection{Double Higgs and Higgs plus jet productions }
The presence of top partners and mixing with top quark has consequences on both double Higgs and double dilaton production. When $\sS{}=0$ and $\sL{}=0$, 
there is no contribution from top partners or dilaton to the production of a pair of Higgs bosons at hadron colliders. When $\sL{}\ne0$, top partners begin to 
contribute to both the triangle and box diagrams in double Higgs production. At the same time, the contribution from the top quark reduces due to smaller values of 
the top Yukawa coupling. Since top partners are more massive than the top itself, the consequence of having $\sL{}\ne 0$ is to reduce the double Higgs production 
cross  section in the invariant mass region that is far below the mass of the $t'$~\cite{Dawson:2015oha}.\footnote{It is interesting to note that since the mass 
eigenstates are rotated through orthogonal transformation, the cross section follows a sum rule, namely that the double Higgs cross section reduces to the SM value when 
$m_{t}=m_{t'}$.} Needless to say that if the mass of $t'$ is not too large and can be directly produced in high energy colliders, a resonance peak in the double Higgs boson 
invariant mass distribution is expected. Interestingly, when $\sL{}\ne 0$, then $s \to hh$ decays are possible, albeit small. This would give rise to a small peak in the invariant mass spectrum at $m_{hh}~\sim 750$ GeV.

Similarly, the presence of the top partners will also affect the transverse momentum ($p_T$) distribution of the Higgs boson produced in association 
with a high $p_T$ jet \cite{Dawson:2015oha}. When $\sL{}\ne0$, top partners begin to contribute to both the triangle and box diagrams in Higgs plus jet production.
In the small  $p_T$ (less than $m_h$) region,  the $\sL{}$ dependence of the differential cross section is weak,\footnote{ 
The reason is the same as we have argued at the end of section~\ref{sec:higgs-fit} when studying the decay width of $h \to gg$ in the MDM, 
since both $m_{t}$ and $m_{t'}$ are larger than $m_h/2$.}
consequently, the low $p_T$ distribution of the Higgs boson will approximately resemble the SM prediction.
However, in the high $p_T$ (larger than $m_t$) region, where the invariant $(\sqrt{\hat{s}})$ of the process is large, the 
differential cross section will depend on the value of $\sL{}$  and become smaller than the SM prediction 
as $p_T$ increases toward $m_{t'}$ . 
Hence, a precise measurement of the $p_T$ distribution, particularly in the high $p_T$ region, can probe 
BSM physics such as the MDM.

\section{Conclusions}
In this work we analyzed the Minimal Dilaton Model as a potential candidate to explain the $750$ GeV diphoton excess. The linearized version of the model presents an effective lagrangian that includes the usual SM along with a singlet scalar dilaton ($s$) and $N_T$ number of vector-like top partners with quantum numbers identical to the right handed top quark. 

We identify the possible $750$ GeV resonance with the dilaton $s$ and find that the model can explain the diphoton signal and avoid other experimental constraints. We find that values of $f$ in the range of $[0.4,1]$~TeV can explain the diphoton excess in this model. 
Interestingly, we find that in order to explain the signal, mixing between the Higgs boson and the dilaton must be very small or zero (i.e., $\sin\theta_S \sim 0)$. Mixing between the top and its partner(s)
 is allowed but is constrained ($\sL{} \leq 0.3$) by precision electroweak test (i.e., $S$, $T$ and $U$ parameters) and by single top production measurements.
We also note that the mixing angle $\sL{}$ may be further constrained through experiments by measuring the
$s\to Z\gamma$,  $s\to Z Z$, and $s\to t\bar{t}$ decay branching ratios,  and by improved single-top measurements. The constraint on $\sin\theta_L$ from single-top measurement is independent of the mass of the partner. This conclusion also holds for any model which allows the top partner to mix with top quark.

We observe three key features with regards to the phenomenology of the MDM. Firstly, signatures for the top partner(s) at the LHC are archetypal of vector like quark searches. 
However, the model allows the presence of additional top partners (to explain the diphoton excess). 
Hence, so long as $\sL{}\ne 0$, one hopes to find non-degenerate top partners at the LHC.
Secondly, double higgs production is affected when $\sL{}\ne0$. Finally, Higgs plus jet production at the LHC in the high $p_T$ region is also sensitive to presence of top partner(s) in the loop.
\section{Acknowledgements}
This work was supported by the National Science Foundation under Grant No. PHY-1417326 and under Grant No. PHY-1519045. We thank S. Dawson and C.-P. Yuan for discussions. 
We also wish to acknowledge the support of the Michigan State University High Performance Computing Center and the
Institute for Cyber Enabled Research. 

\appendix
\section{Decays of $h$}
\label{app:hdecay}
We identify $h$ with the $125$ GeV Higgs boson. The decay channels we consider are listed below.
\begin{equation}
\left\{\Gamma(h \to f\bar{f}),
\Gamma(h \to ZZ),
\Gamma(h \to W^{+} W^{-}),
\Gamma(h \to \gamma \gamma),
\Gamma(h \to Z \gamma),
\Gamma(h \to g g)
\right\}
\end{equation}
Here, the fermions considered are:
\begin{equation}
f=\left\{
s,c,b,\tau,\mu
\right\}.
\end{equation}
Note that the last three decay channels are loop induced and we will consider the ${b,t,t'}$ fermions in the loop process as well as $W$ gauge bosons in the loop.
The decay widths of $h$ scale according to the couplings in Eq.~(\ref{eqn:s-coup}). We therefore write below the ratio of $h$ decay widths to the $h_{SM}$ decay widths.
\begin{equation}
R_{hff}=\frac{\Gamma(h\to f\bar{f})}{\Gamma(h_{SM}\to f\bar{f})} = C_{hff}^2,
\end{equation}
\begin{equation}
R_{hVV}=\frac{\Gamma(h\to VV)}{\Gamma(h_{SM}\to VV)} = C_{hVV}^2.
\end{equation}

\subsection{Decay of $h\to \gamma \gamma$}
The partial decay width for $h\to \gamma \gamma$ is~\cite{Djouadi:1996yq}  
\begin{gather}
\Gamma(h_{SM} \to \gamma \gamma)= \frac{G_{\mu} \alpha^2 m_{h}^3}{128 \sqrt{2} \pi^3}
\left|
\sum_{f = b,t} Q_f^2 N_c A_{1/2}(\tau_f) +  A_{1}(\tau_W)
\right|^2,
\end{gather}
where, $A_j$ corresponds to the form factors defined below~\cite{Djouadi:2005gj}:
\begin{eqnarray}
A_{1/2}(\tau)&=&\frac{2}{\tau^2}(\tau + (\tau -1)f(\tau)), \\ \nn
A_{1}(\tau)&=&-\frac{1}{\tau^2}(2\tau^2 +3\tau +3(2\tau -1)f(\tau)),
\label{eq:app:A}
\end{eqnarray}
with $\tau_i=\frac{m_h^2}{4m_i^2}$ and
\begin{eqnarray}
f (\tau) = \left\{ \begin{array}{lc}
\mbox{arcsin}^2 \sqrt{\tau}  & \mbox{for }\tau \leq 1 \\
- \frac{1}{4} \left[ \log \frac{1+\sqrt{1-\tau^{-1}}}{1-\sqrt{1-\tau^{-1}}} - i \pi \right]^2 & 
\mbox{for }\tau > 1 \end{array} \right.\,.
\end{eqnarray}

\begin{equation}
R_{h\gamma\gamma}=\frac{\Gamma(h\to \gamma \gamma)}{\Gamma(h_{SM}\to \gamma \gamma)} = \frac{\left| C_{hWW}A_1(m_h^2/4m_W^2) + \sum_{f = b,t,t'}C_{hff}A_{1/2}(m_h^2/4m_f^2) \right|^2}{\left|A_1(m_h^2/4m_W^2) + \sum_{f = b,t}A_{1/2}(m_h^2/4m_f^2) \right|^2}.
\end{equation}

\subsection{Decay of $h\to gg$}
The partial decay width for $h\to gg$ is~\cite{Djouadi:1996yq}  
\begin{gather}
\Gamma(h_{SM} \to gg)= \frac{G_{\mu} \alpha_s^2 m_{h}^3}{36 \sqrt{2} \pi^3}
\left|
\sum_{f = b,t} \frac{3}{4} A_{1/2}(\tau_f)
\right|^2,
\end{gather}
where, $A_{1/2}$ is given in Eq.~\ref{eq:app:A}. In the MDM the decay width is determined as follows:
\begin{equation}
R_{hgg}=\frac{\Gamma(h\to gg)}{\Gamma(h_{SM}\to gg)} = \frac{\left| \sum_{f = b,t,t'}C_{hff}A_{1/2}(m_h^2/4m_f^2) \right|^2}{\left| \sum_{f = b,t}A_{1/2}(m_h^2/4m_f^2) \right|^2}.
\end{equation}
\subsection{Decay of  $h\to Z \gamma$}
The partial decay width for $h\to Z \gamma$ is ~\cite{Djouadi:1996yq}
\begin{gather}
\Gamma(h_{SM} \to Z \gamma)= \frac{G_{\mu}^2 M_W^2\alpha m_{h}^3}{64 \pi^4}\left(1 - \frac{M_Z^2}{m_h^2}\right)^3 \\\nonumber \times
\left|
\sum_{f = b,t} Q_f N_c \frac{ 2I_3^f -4s_W^2 Q_f}{c_W}\mathcal{A}_{f}(\tau_f,\lambda_f) +  \mathcal{A}_{W}(\tau_W,\lambda_W)
\right|^2,
\end{gather}
where,
\begin{eqnarray}
\mathcal{A}_f(\tau_f,\lambda_f) &=&   I_1(\tau_f,\lambda_f) - I_2(\tau,\lambda)\nonumber, \\
\mathcal{A}_W(\tau_W,\lambda_W) &=& c_W \left[ \left(1+\frac{2}{\tau_W}\right)\frac{s_W^2}{c_W^2}   -  \left(5+\frac{2}{\tau_W}\right) 
\right] I_1(\tau_W,\lambda_W) \nonumber \\
&+&  4 c_W \left( 3 - \frac{s_W^2}{c_W^2} \right) I_2(\tau_W,\lambda_W),
\end{eqnarray}
\begin{eqnarray}
I_1(\tau,\lambda) &=& \frac{\tau \lambda}{2(\tau-\lambda)}
+\frac{\tau^2 \lambda}{2(\tau-\lambda)^2}
\Big( \lambda \left[f(\tau)-f(\lambda)\right] 
+ 2 \left[g(\tau)-g(\lambda)\right] \Big),
\nonumber \\
I_2(\tau,\lambda) &=&  -\frac{\tau \lambda}{2(\tau-\lambda)}
\left[f(\tau)-f(\lambda)\right] ,
\end{eqnarray}
and  $\lambda=4m^2/M_Z^2$ and $\tau=4m^2/m_{h}^2$.
with the functions $f$ and $g$ defined by 
\begin{equation}
f(\tau) = \left\{ \begin{array}{ll}
{\rm arcsin}^2 \sqrt{1/\tau} & \tau \geq 1 \\
-\frac{1}{4} \left[ \log \frac{1 + \sqrt{1-\tau } }
{1 - \sqrt{1-\tau} } - i \pi \right]^2 \ \ \ & \tau <1
\end{array} \right. ,
\end{equation}
\begin{equation}
g(\tau) = \left\{ \begin{array}{ll}
\sqrt{\tau-1} \ {\rm arcsin} \sqrt{1/\tau} & \tau \geq 1 \\
\frac{1}{2} \sqrt{1-\tau} \left[ \log \frac{1 + \sqrt{1-\tau } }
{1 - \sqrt{1-\tau} } - i \pi \right] \ \ \ & \tau <1
\end{array} \right. ,
\end{equation}

\begin{equation}
R_{hZ\gamma}=\frac{\Gamma(h\to Z \gamma)}{\Gamma(h_{SM}\to Z \gamma)} = 
\frac{
\left|
\sum_{f = b,t,t'}C_{hff} Q_f N_c \frac{ 2I_3^f -4s_W^2 Q_f}{c_W}\mathcal{A}_{f}(\tau_f,\lambda_f) +  C_{hWW}\mathcal{A}_{W}(\tau_W,\lambda_W)
\right|^2
	}
	{\left|
		\sum_{f = b,t} Q_f N_c \frac{ 2I_3^f -4s_W^2 Q_f}{c_W}\mathcal{A}_{f}(\tau_f,\lambda_f) +  \mathcal{A}_{W}(\tau_W,\lambda_W)
		\right|^2} .
\end{equation}
Note that in the expressions above we keep only the leading terms and  have neglected diagrams that contribute to decay amplitude as $\cL{2}\sL{2}$.
\section{Decays of the dilaton s}
\label{app:sdecay}
We identify $s$ with the $750$ GeV excess. The decay channels we consider are listed below.
\begin{equation}
\left\{\Gamma(s \to f\bar{f}),
\Gamma(s \to hh),
\Gamma(s \to ZZ),
\Gamma(s \to W^{+} W^{-}),
\Gamma(s \to \gamma \gamma),
\Gamma(s \to Z \gamma),
\Gamma(s \to g g)
\right\}
\end{equation}
Here,
\begin{equation}
f=\left\{
s,c,b,t,\tau,\mu
\right\}.
\end{equation}
The relations are similar to the Higgs ratios written down in the section above, except for the  $s \to ZZ$ decays where we need to consider the $s\to ZZ$ loop induced decay.

\subsection{Decays to fermions }
\begin{equation}
\Gamma(s\to f\bar{f}) =C_{sff}^2\frac{G_{\mu} N_c}{4\sqrt{2}\pi} m_{s} m_{f}^2 \beta_{f}^3 ,
\end{equation}
where $\beta_f = (1 - 4m_f^2/m_{s}^2)^{1/2}$.

\subsection{ Decays of $s \to VV$}
Here we consider only the on-shell decay of the massive gauge bosons
\begin{equation}
\Gamma(s \to VV) = \frac{G_{\mu} m_{s}^3}{16 \sqrt{2}\pi}\delta_{V}(x)\ ;\ x=\frac{M_V^2}{m_s^2}
\end{equation}
Here $\delta_{W} =2 C_{sWW}^2 \sqrt{1 -4x}(1 -4x +12 x^2)$ and 
\begin{equation}
	\delta_{Z}(x)=  (C_{sZZ}^2 \sqrt{1 -4x}(1 -4x +12 x^2) + \frac{C^2_{st't'}}{m_{t'}^2}\frac{\alpha M_W^2 s_W^2 N_c^2}{18 \pi^2 c_W^4 s_W^4}(2 s_W^4 (2/3)^2)^2 \sqrt{1 -4x}(1 -4x +6 x^2))
\end{equation}
The second term is the contribution from a massive $t'$ loop to the $s\to ZZ$ process. This has been calculated using the infinitely massive $t'$ approximation 
in a low energy theorem~\cite{Shifman:1979eb}. We keep only the leading terms and do not include top quarks to the loops as these are suppressed by powers of $\sL{}$.

\subsection{ Decays of $s \to \gamma \gamma$}
Similar to the Higgs decays, the dilaton decay is given as 
\begin{gather}
\Gamma(s \to \gamma \gamma)= \frac{G_{\mu} \alpha^2 m_{h}^3}{128 \sqrt{2} \pi^3}
\left|
\sum_{f = b,t,t'} C_{sff}Q_f^2 N_c A_{1/2}(\tau_b) +  C_{sWW}A_{1}(\tau_W)
\right|^2
\end{gather}

\subsection{ Decays of $s \to gg$}
Similar to the Higgs decays, the dilaton decay is given as 
\begin{gather}
\Gamma(s \to \gamma \gamma)= \frac{G_{\mu} \alpha_s^2 m_{h}^3}{36 \sqrt{2} \pi^3}
\left|
\frac{3}{4}\sum_{f = b,t,t'} C_{sff} A_{1/2}(\tau_b) 
\right|^2
\end{gather}

\subsection{ Decays of $s \to Z \gamma$}
Similar to the Higgs decays the dilaton decay is given as 
\begin{gather}
\Gamma(s \to Z \gamma)= \frac{G_{\mu}^2 M_W^2\alpha m_{s}^3}{64 \pi^4}\left(1 - \frac{M_Z^2}{m_s^2}\right)^3 \\\nonumber \times
\left|
\sum_{f = b,t,t'} C_{sff}Q_f N_c \frac{ 2I_3^f -4s_W^2 Q_f}{c_W}\mathcal{A}_{f}(\tau_f,\lambda_f) +  C_{sWW}\mathcal{A}_{W}(\tau_W,\lambda_W)
\right|^2 .
\end{gather}

\subsection{Decays of $s \to h h$}      
The scalar potential is 
\begin{equation}
V(S,H)= \frac{m_S^2}{2} S^2 + \frac{\lambda_S}{4}S^4 + \frac{\kappa}{2}S^2 |H|^2 + m_{H}^2 |H|^2 + \frac{\lambda_H}{4}|H|^4 .
\end{equation}
Note that the couplings $\kappa,\lambda_S,\lambda_H$ can be written in terms of the paramters $(\theta_S,f,v, m_h,m_s)$. Where $m_h$ and $m_s$ are the physical masses,

\begin{eqnarray}
\kappa &=& \frac{(m_s^2 - m_h^2) \cS{}\sS{}}{\sqrt{2} f v}  , \nonumber\\
\lambda_H &=& \frac{m_h^2\cS{2} + m_s^2\sS{2}}{v^2},
\nonumber\\
\lambda_S &=& \frac{m_h^2\sS{2} + m_s^2\cS{2}}{2f^2}.
\label{lambda} 
\end{eqnarray}

From the expressions above, we derive the $shh$ vertex as follows
\begin{align*}
C_{shh}=\frac{-i\sS{} \cS{} }{8 f v}
&\bigg[
f \sS{} \left(\left(\sqrt{2}-3\right) m_h^2-\left(9+\sqrt{2}\right) m_s^2\right)-3 \left(1+\sqrt{2}\right) f \sin (3 \theta_S ) (m_h^2-m_s^2)&\nonumber \\
&+v\cS{} \left(\left(\sqrt{2}-6\right) m_h^2-\left(18+\sqrt{2}\right) m_s^2\right)+3 \left(2+\sqrt{2}\right) v \cos (3 \theta_S ) (m_h^2-m_s^2)
\bigg]&
\end{align*}

The paritla decay width of $ s \to h h$ can be written as follows:
\begin{equation}
\Gamma(s \to h h) = \frac{C_{shh}^2}{32\pi m_s}\left(1 - \frac{4m_h^2}{m_s^2}\right)^{1/2} .
\end{equation}

\subsection{Decay Branching Ratios of the Dilaton}
\label{app:bratio}
We present here variation of the branching ratios of the dilaton $s$ with various parameters of the model.

\begin{figure}[ht]
\centering
\begin{minipage}{0.45\textwidth}
	\includegraphics[scale=0.5]{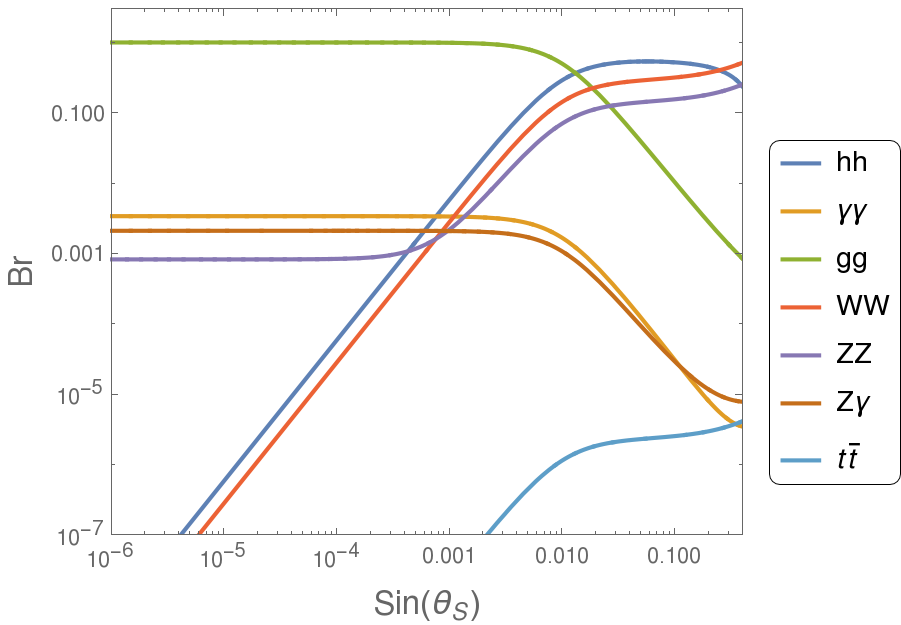}
	\caption{Branching ratios of s for $\sL{}=0$, and $\frac{v}{f}=1$. \label{fig:app:BR-s}}
\end{minipage}
\hfill
\begin{minipage}{0.45\textwidth}
	\includegraphics[scale=0.5]{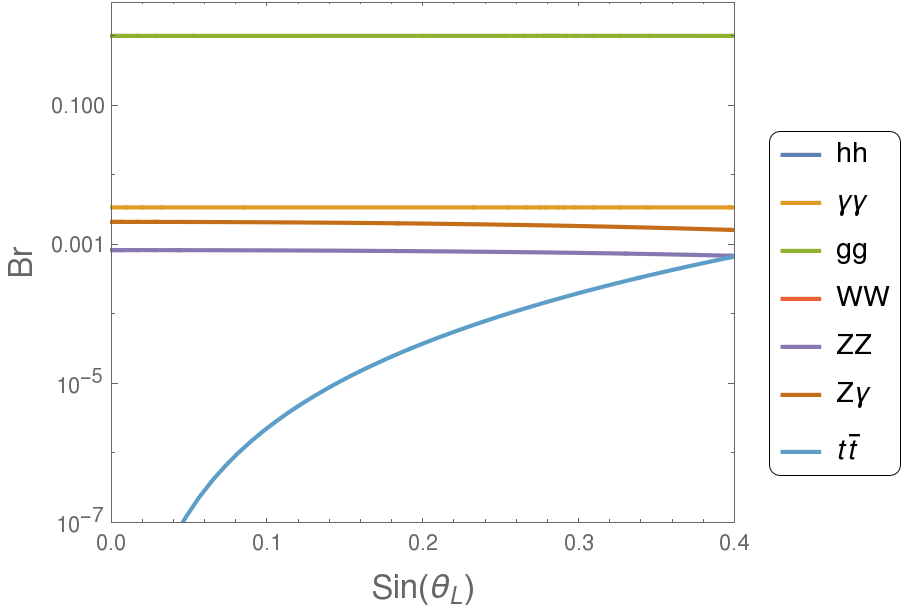}
	\caption{Branching ratios of s for $\sS{}=0$, and $\frac{v}{f}=1$.\label{fig:app:BR-l}}
\end{minipage}
\end{figure}

\section{Top and top partner mass matrix}
\label{app:massmat}
\subsection{$N_T=1$}
The mass matrix can be written down below:
	
	\al{
		\bb
		\oli{q_{3L}} & \oli{T_L}
		\eb
		\bb
		m	&	m'\\
		0	&	M
		\eb
		\bb
		u_{3R} \\ T_R
		\eb,
		\label{our choice}
	}
	where $m=y_tv/\sqrt{2}$ and $m'=y'v/\sqrt{2}$. Switching to mass eigenstates
	\al{
		\bb
		q_{3L}\\
		T_L
		\eb
		&=	\bb
		\cos\theta_L	&	\sin\theta_L\\
		-\sin\theta_L	&	\cos\theta_L
		\eb
		\bb
		t_L\\
		t'_L
		\eb,	&
		\bb
		u_{3R}\\
		T_R
		\eb
		&=	\bb
		\cos\vartheta_R	&	\sin\vartheta_R\\
		-\sin\vartheta_R	&	\cos\vartheta_R
		\eb
		\bb
		t_R\\
		t'_R
		\eb,
	}
	we may diagonalize as
	\al{
		\bb
		\oli{q_{3L}} & \oli{T_L}
		\eb
		\bb
		m	&	m'\\
		0	&	M
		\eb
		\bb
		u_{3R} \\ T_R
		\eb
		&=	\bb
		\oli{t_L} & \oli{t'_L}
		\eb
		\bb
		m_t	&	0\\
		0	&	m_{t'}
		\eb
		\bb
		t_R	\\ t'_R
		\eb,
	}
	where
	\al{
		\tan\theta_L
		&=	{\sqrt{\paren{M^2-m^2+m'^2}^2+4m'^2m^2}-M^2+m^2+m'^2\over2m'M}
		=	{m'\over M}+O(M^{-3}),	\nn
		\tan\vartheta_R
		&=	{\sqrt{\paren{M^2-m^2+m'^2}^2+4m'^2m^2}-M^2+m^2-m'^2\over2m'm}
		=	{m'm\over M^2}+O(M^{-4}),
	}
	and the mass eigenvalues are
	\al{
		\left\{\begin{array}{c}m_t^2\\ m_{t'}^2\end{array}\right\}
		&=	
		{M^2+m^2+m'^2\mp\sqrt{\paren{M^2+m^2+m'^2}^2-4m^2M^2}\over2}.
	}
	For large $M$,
	\al{
		m_t
		&=	\paren{1-{m'^2\over2M^2}}m+O(M^{-4}),\nn
		m_{t'}
		&=	M+{m'^2\over2M}+O(M^{-3}).
	}
	Conversely, parameters in the Lagrangian can be written in terms of the observables:
	\begin{align}
	M
	&=
	\sqrt{ m_t^2 \sin^2 \theta_L + m_{t'}^2 \cos^2 \theta_L}
	, \\ 
	y_t
	&=
	\frac{\sqrt{2}}{v}
	\frac{m_t m_{t'}}{\sqrt{ m_t^2 \sin^2 \theta_L + m_{t'}^2 \cos^2
			\theta_L}}
	, \\ 
	y'
	&=
	\frac{\sqrt{2}}{v}
	\frac{(m_{t'}^2 - m_t^2) \sin \theta_L \cos \theta_L}{\sqrt{ m_t^2
			\sin^2 \theta_L + m_{t'}^2 \cos^2 \theta_L}} 
	.
	\end{align}

\subsection{$N_T>1$}
The mass matrix can be written down below:
	
	\al{
		\bb
		\oli{q_{3L}} & \oli{T_L} & \oli{T_{2L}} & ...
		\eb
		\bb
		m	&	m' & m' & ... & m'\\
		0	&	M & 0  & ... & 0 \\
		0       &       0 & M  & ... & 0 \\
		\vdots & \vdots& \vdots  &  \ddots & 0\\
		0	&	0  & 0 & ... & M
		\eb
		\bb
		u_{3R} \\ T_R \\ T_{2R} \\ \vdots
		\eb,
		\label{ourChoice2}
	}
where the above matrix has dimensions $(N_T+1) \times (N_T+1)$. The diagonal mass squared matrix is given by:
	
	\al{
		\bb
		m^2_t	&	0 & 0 &  ... & 0\\
		0	&	m^2_{t'} & 0  & ... & 0 \\
		0       &       0 & m^2_{t'_2} & ... & 0 \\
		\vdots & \vdots& \vdots  &  \ddots & 0\\
		0	&	0  & 0 & ... & m^2_{t'_{N_T}}
		\eb,
	}

Assumming all the mixings are small and approximately equivalent, we can rotate the diagonal mass squared matrix back to the non-diagonal matrix given by:

	\al{
	    \mathcal{MM}^T = \bb
		m^2 + N_T m'^2	&	M m' & M m' &  ... & M m'\\
		M m'	&	M^2 & 0  & ... & 0 \\
		M m'       &       0 & M^2 & ... & 0 \\
		\vdots & \vdots& \vdots  &  \ddots & 0\\
		M m'	&	0  & 0 & ... & M^2
		\eb .
	}
	\al{\mathcal{MM}^T
	    = \bb
		m_t^2 	& (m^2_t - m^2_{t'}) \theta_L & (m^2_t - m^2_{t'_2}) \theta_L & ... & (m^2_t - m^2_{t'_{N_T}}) \theta_L \\
		(m^2_t - m^2_{t'}) \theta_L & m^2_{t'}	& 0 & ... & 0 \\
		(m^2_t - m^2_{t'_2}) \theta_L & 0 & m^2_{t'_2} & ... & 0\\
	        \vdots & \vdots & \vdots & \ddots & 0 \\
		(m^2_t - m^2_{t'_{N_T}}) \theta_L & 0 & 0 & ... & m^2_{t'_{N_T}} 
	      \eb + \mathcal{O}(\theta_L^2). 
	}

The eigenvalues in the small mixing limit are given by:
	\al{
	    m^2_t \approx m^2-\frac{N_T m^2 m'^2}{M^2}\ , \quad 
	    m^2_{t'} \approx M^2 + N_T m'\ ,\quad 
	    m^2_{t'_2} = ... = m^2_{t'_{N_T}} = M^2\ .
	    \label{eq:app-mass-eval}
	}

The degeneracy of the new particles is broken slightly by the mixing, but only occurs for one of them. This results in a particle slightly heavier than the other new 
top particles.

The rotation matrix to rotate from the flavor eigenstates to the mass eigenstates is given by a product of rotations matrices, $R_1$, $R_2$, ..., $R_{N_T}$. The first one
is given here as a reference, and the rest are related to it by shifting which componenents get mixed with the $t$.

	\al{ 
	      \bb 
		\cos(\theta_L) & \sin(\theta_L) & 0 & ... & 0 \\
		-\sin(\theta_L) & \cos(\theta_L) & 0 & ... & 0 \\
		 0		&	0	 & 1 & ... & 0 \\
		\vdots		& \vdots	& \vdots & \ddots & 0 \\
		0		&	0	& 0	& ... & 1 
	      \eb.
	}

Modified couplings for the case when $N_T>1$ (assuming  $\sin(\theta_S)=0$)  are given below. These couplings are valid only in the small mixing limit $\sL{}\ll 1$.

For the Higgs field:
\begin{equation}
\tilde{C}_{htt}=\frac{m_t}{v}\cL{2{N_T}}\ ,\quad 
\tilde{C}_{ht{t'_i}}=\frac{m_{t'}}{v} \sL{}\cL{i-1+{N_T}}\ , \quad 
\tilde{C}_{h{t'_i}{t'_j}}=\frac{m_{t'}}{v}\sL{2}\cL{i+j-2}\ .
\end{equation}

For the Dilaton field:
\begin{equation}
\tilde{C}_{stt}=\frac{m_t}{v} \eta (1-\cL{2{N_T}})\ ,\quad
\tilde{C}_{st{t'_i}}=\frac{m_{t'}}{v} \eta (- \sL{}\cL{i-1+{N_T}})\ , \quad
\tilde{C}_{s{t'_i}{t'_j}}=\frac{m_{t'}}{v} \eta (\delta_{i j} - \sL{2} \cL{i+j-2})\ .
\end{equation}

For the Z boson:
\begin{align}
&\tilde{C}_{Ztt}=\frac{g}{2 \cos(\theta_W)} (\cL{2{N_T}} {P_L} - 2Q \sin^2 \theta_W )\ ,\quad
\tilde{C}_{Zt{t'_i}}=\frac{g}{2 \cos(\theta_W)} (\sL{} \cL{i-1+{N_T}} {P_L})\ ,&\nn 
&\tilde{C}_{Z{t'_i}{t'_j}}=\frac{g}{2 \cos(\theta_W)} ((\sL{2} \cL{i+j-2} {P_L} - 2 \delta_{i j} Q \sin^2 \theta_W )\ .&
\end{align}

For W-bosons:
\begin{equation}
\tilde{C}_{Wtb}=\frac{g}{2 \sqrt{2}}  (\cL{{N_T}} {P_L} )\ ,\quad
\tilde{C}_{W{t'_i}b}=\frac{g}{2 \sqrt{2}}  ({P_L} \sL{} \cL{i-1})\ ,\\
\end{equation}
where $i,j = \{1 , 2 ,.... , N_T\}$.

\section{Vaccuum Stability}
\label{app:vac-stab}

To check the validity of the theory, the scale at which the effective theory breaks down is calculated. This is done by considering the runnning of the coupling
constants. This is traditionally done using the renormalization group equation defined as:
\begin{equation}
\beta = \frac{dg}{d\log(\mu)} = \frac{1}{16\pi^2}\beta^{(1)}+\frac{1}{(16\pi^2)^2}\beta^{(2)}+... \ ,
\end{equation}
where $\beta^{(1)}$ and $\beta^{(2)}$ are the one- and two-loop contributions to the beta function, respectively, and $\mu$ is the renormalization scale. For simplicity, only the one
loop corrections are considered for this analysis. For the gauge couplings, the $\beta$ functions are given by~\cite{1512.08307,1502.01361,1404.0681}:
\begin{equation}
\beta^{(1)}_{g_{1}}=\frac{16}{9} N_{T} g_{1}^{3} + \frac{41}{6} g_{1}^{3}   ,\quad
\beta^{(1)}_{g_2}=-\frac{19}{6}g_2^3 ,\quad
\beta^{(1)}_{g_3}=\left(-7+N_{T}\frac{2}{3}\right)g_3^3.
\end{equation}

The Yukawa coupling sector has $\beta$ functions given by:

\begin{eqnarray}
\beta_{y_{t}}&=&
\frac{1}{16 \pi^2} y_t (-\frac{17}{12} g_1^2 - \frac{9}{4} g_2^2 - 8 g_3^2 + \frac{9}{2}N_T y'^2 + \frac{9}{2} y_t^2)\nonumber \\
\beta_{y_{x}}&=&\frac{1}{16 \pi^2}y_x(
-\frac{8}{3} g_1^2  -8 g_3^2 +N_T y'^2 + 3 (1 + 2 N_t) y_x^2) \nonumber \\
\beta_{y'}&=&\frac{1}{16 \pi^2} y'
(-\frac{17}{12} g_1^2  - \frac{9}{4} g_2^2  - 8 g_3^2  + \frac{9}{2} N_T y'^2 + \frac{9}{2}
y_t^2 + \frac{1}{2} y_x^2)
\end{eqnarray}
where $y_T = \frac{M}{f}\cos(\theta_L)$ at the scale of the $t'$ mass. Finally, the scalar sector $\beta$ functions are:
\begin{align}
\beta_{\lambda_{H}}=\frac{1}{16 \pi^2}
&\bigg[
\frac{3}{2} g_1^4 + 3 g_1^2 g_2^2 + \frac{9}{2} g_2^4 + 2 \kappa^2 - 
3 g_1^2 \lambda_H
- 9 g_2^2 \lambda_H + 6 \lambda_H^2 & \nonumber \\
&+ 12 N_T \lambda_H y'^2 - 
24 N_T y'^4 + 12 \lambda_H  y'^2
+ 12 \lambda_H  y_t^2 - 48 N_T y'^2 y_t^2 - 24 y_t^4
\bigg]&
\end{align}
\begin{align}
\beta_{\lambda_{S}}=\frac{1}{16 \pi^2}(
2\kappa^2 + 18 \lambda_S^2 + 24 \lambda_S N_T y_x^2 - 24 
N_T y_T^4
)
\end{align}

Here, the coupling between $H$ and $S$ is zero since the mixing between the two is set to zero. Therefore, the running of this coupling is not included. For the determination
of the vaccuum stability, the limit of small mixing will be taken, and the coupling $y'$ will be taken to be zero, for simiplicity. The scale that is used to
set the parameters is the $M_Z$ scale for all the parameters, with the exception of $\lambda_S$, whose values are zero until the scale reaches the new physics scale determined by $m_{t'}$ and $m_s$ respectively.
\begin{gather}
g_1\left(\mu=M_Z\right) = \frac{\sqrt{4\pi\alpha(M_Z)}}{\sqrt{1-\sin(\theta_W)^2}},\ \ \ \ 
g_2\left(\mu=M_Z\right) = \sqrt{\frac{8 G_F}{\sqrt{2}}}M_W, \nonumber \\ 
g_3\left(\mu=M_Z\right) = \sqrt{4\pi\alpha_s(M_Z)}, \ \ \ \ 
\lambda_H\left(\mu=M_Z\right) = \frac{1}{\sqrt{2}}G_F M_H^2(1+\delta_H), \quad
y_t\left(\mu=M_Z\right) = \sqrt{\sqrt{2}G_F} m_t (1+\delta_t),
\end{gather}
where $\alpha(M_Z)$ is the fine structure constant at the scale of $M_Z$, $G_F$ is the Fermi constant, $M_W$ the mass of the W boson, $\alpha_s(M_Z)$ the strong coupling at
the scale of $M_Z$, and $\delta_H$ and $\delta_t$ are the one-loop corrections to the quartic coupling of the Higgs and the Yukawa coupling of the top, respectively. These
corrections are given in Refs.~\cite{SIRLIN1986389,hep-ph/9408313}. Finally, the starting value for $\lambda_S(M_S)$ is $\frac{M_S^2}{2f^2}$ when it is introduced
at the scale of the dilaton mass ($M_S$).

Since relatively large values of $v/f$ are needed to explain the diphoton signal. This implies large values of the top partner Yukawa $y_T$, which contribute negatively to $\beta_{\lambda_S}$ and will make the vacuum unstable. We find that $\lambda_S$ becomes negative rapidly after reaching the top partner mass scale and the theory breaks down. Since the MDM is an effective theory in itself this indicates the presence of further new physics in the TeV range that stabilizes the vacuum. It is possible that the additional dynamics stabilizing the vacuum does not affect the diphoton rates. Hence the MDM can explain the diphoton signal however it requires the presence of additional TeV range physics.

\clearpage
\bibliographystyle{utphys}
\bibliography{mdm_draft_v6}
\end{document}